\documentclass[12pt, draftclsnofoot, onecolumn]{IEEEtran}
\usepackage{lipsum}
\usepackage{cite}
\usepackage{amsmath,amssymb,amsfonts}
\usepackage{algorithmic}
\usepackage{graphicx}
\usepackage{textcomp}
\usepackage{xcolor}
\def\BibTeX{{\rm B\kern-.05em{\sc i\kern-.025em b}\kern-.08em
    T\kern-.1667em\lower.7ex\hbox{E}\kern-.125emX}}
    
\usepackage{subfig}
\usepackage[utf8]{inputenc}
\usepackage{indentfirst}
\usepackage{endnotes}
\usepackage{amsmath}
\usepackage{float}
\usepackage{booktabs}
\usepackage{tabularx}
\usepackage{graphics}
\usepackage{xspace}
\graphicspath{{./images/}}
\usepackage{svg}
\usepackage{pifont}
\usepackage{caption}
\usepackage{todonotes}
\usepackage{listings}
\usepackage{marvosym}
\usepackage{hyperref}
\usepackage{multirow}
\usepackage{todonotes}
\usepackage{colortbl}

\newcommand \ourwork{\textsf{PORTulator}\xspace}
\newcommand {\bsub}[1]{\vspace{3pt}\noindent\textbf{#1}}
\newcommand \modify[1]{\textcolor{black}{#1}}

\title{Physical-Layer Signal Injection Attacks on EV Charging Ports: Bypassing Authentication via Electrical-Level Exploits}
\author{
  Hetian Shi$^{1}$, Yi He$^{1}$, Shangru Song$^{1}$, Jianwei Zhuge$^{1}$, Jian Mao$^{2}$\\
  $^{1}$Tsinghua University, Beijing, China\\
  $^{2}$Beihang University, Beijing, China\\
  Email: shiht18@tsinghua.org.cn, clangllvm@126.com, ssr22@mails.tsinghua.edu.cn, zhugejw@tsinghua.edu.cn, maojian@buaa.edu.cn
}

\begin{document}

\maketitle

%%%% 6. ABSTRACT %%%%

\begin{abstract}
The proliferation of electric vehicles in recent years has significantly expanded the charging infrastructure while introducing new security risks to both vehicles and chargers.
In this paper, we investigate the security of major charging protocols such as SAE J1772, CCS, IEC 61851, GB/T 20234, and NACS, uncovering new physical signal spoofing attacks in their authentication mechanisms.
By inserting a compact malicious device into the charger connector, attackers can inject fraudulent signals to sabotage the charging process, leading to denial of service, vehicle-induced charger lockout, and damage to the chargers or the vehicle's charge management system.
To demonstrate the feasibility of our attacks, we propose \ourwork, a proof-of-concept (PoC) attack hardware, including a charger gun plugin device for injecting physical signals and a wireless controller for remote manipulation.
By evaluating \ourwork on multiple real-world chargers, we identify 7 charging standards used by 20 charger piles that are vulnerable to our attacks.
The root cause is that chargers use simple physical signals for authentication and control, making them easily spoofed by attackers.
To address this issue, we propose enhancing authentication circuits by integrating non-resistive memory components and utilizing dynamic high-frequency Pulse Width Modulation (PWM) signals to counter such physical signal spoofing attacks.

\end{abstract}

%%%% 5. KEYWORDS %%%%
% \keywords{Charging piles \and Charging ports \and Spoofing signal \and Hardware Reverse Engineering}

%%%% 7. PAPER CONTENT %%%%
\section{Introduction} \label{Sec:intro}

The widespread adoption of EVs represents a transformative shift towards sustainable transportation, addressing environmental concerns and reducing dependence on fossil fuels.
Meanwhile, new security issues~\cite{iehira2018spoofing,conti2022evexchange,kailus2024self} are emerging within the EV charging infrastructure.
Unfortunately, as more EV chargers are deployed in cities, they become increasingly attractive targets for cyberattacks. 
Moreover, the complexity and diverse range of charging standards open the door to numerous vulnerabilities that could threaten both user safety and the stability of critical charging infrastructures~\cite{venvckauskas2024enhancing}.

Existing works~\cite{kohler2022brokenwire, baker2019losing, nasr2023chargeprint} mostly focus on remote attacks. For instance, the Brokenwire attack~\cite{kohler2022brokenwire} can disrupt the Combined Charging System (CCS) charging process by performing remote electromagnetic interference on the charger's programmable logic controller (PLC) to terminate the charging session.
\modify{While Nasr et al.~\cite{nasr2023chargeprint} and Vailoces et al.~\cite{vailoces2023securing} both analyze vulnerabilities in the backend of electric vehicle supply equipment (EVSE) systems, including charge management platforms and network-level authentication, little attention has been paid to the physical-layer signaling protocols and port-level logic at the EV charger interface.}
%Nasr~\cite{nasr2023chargeprint} et al. analyze the attack vectors and measure the vulnerabilities in charge management websites.
%Vailoces et al.~\cite{vailoces2023securing} provided a broad analysis of vulnerabilities in electric vehicle supply equipment (EVSE) systems, including weak backend authentication and insecure end-to-end communication. Their work outlines several attack scenarios and corresponding countermeasures, primarily at the backend server and network level. 

In this work, we investigate local attacks on charger piles and identify new common physical attack vectors that can exploit the weaknesses in the authentication process of several major charging standards. 
We demonstrate practicable physical signal injection attacks on various real-world charger pipes.
By planting a small concealed hardware plugin onto the charging guns, attackers can inject control signals into the various ports of the guns.
Specifically, by manipulating different ports with specific physical signals, attackers can launch: (1) Denial-of-service (DoS) attacks disrupting charging via Charging Confirmation (CC) / Control Pilot (CP) port manipulation; (2) Deadlock attacks spoofing impedance on the CC port, which can lockout the charging gun; and \modify{(3) PWM/CAN Bus signal injection attacks that can damage the EV battery, overloading the charging system, or inject malicious CAN Bus message to further exploiting the inner systems of vehicles.}
We prototype an attack hardware called \ourwork, which can be seamlessly integrated into the charger guns' ports and perform physical signal injection attacks.
\modify{Unlike~\cite{song2023ransom}, which only showed a simple demo of authentication issues, our work is the first to fully study and exploit state forgery problems in several EV charging protocols. We show that attackers can tamper with the physical authentication process and then send fake CAN messages to the vehicle, which can bypass battery safety protections and cause overcharging.} %\todo{R.1 Highlighted our novel contributions beyond [Ano23] and weak-authentication works.} 
%We test this attack across seven common charging standards and confirm that all are affected. We also propose fixes that make it harder to fake these signals, covering both protocol changes and hardware improvements.
We evaluate \ourwork on several real-world chargers and identify that 7 charging standards used by 20 charger piles are vulnerable to our attacks.
Finally, we propose defensive strategies to mitigate these vulnerabilities, offering solutions that enhance the cybersecurity of EVSE systems. 
By integrating non-resistive memory components and utilizing dynamic high-frequency Pulse Width Modulation (PWM) power in existing charging authentication processes, the impedance is changed from fixed values to changeable values that are infeasible to be forged by attackers.
Experiment results show that our solution can effectively counter the physical signals of spoofed attacks.

Here are the contributions of our work:
\begin{itemize}
    \item We first identify critical vulnerabilities in the authentication mechanisms of various EV charging standards, allowing attackers to execute a charging process DoS attack, lock the charging port, manipulate discharge power, and potentially gain unauthorized access to internal systems such as the CAN Bus, posing significant security risks.

    \item We develop \ourwork, an attack suite that integrates a microcontroller unit (MCU), flexible printed circuit (FPC), and wireless communication, enabling covert connection to different charging standards and remote manipulation of vehicle charging port states for various attacks.

    \item We test \ourwork against multiple mainstream charging gun standards, demonstrating its practical effectiveness in real-world scenarios and providing three comprehensive case studies to highlight significant threats to vehicle charging safety.

    \item To address these vulnerabilities, we propose a defensive mechanism and validate it with real-world prototypes. 
    Experimental results show that it can effectively prevent physical signal injection attacks.
    
    % \item In response to these vulnerabilities, we propose and validate defensive strategies that effectively mitigate the identified risks, emphasizing the urgent need for enhanced security measures in EV charging infrastructure.

\end{itemize}

 % 主要侧重点集中于：充电桩与充电桩存在协议认证漏洞
 % 充电口死锁问题（伪造阻抗大小）、充电过程中的DoS攻击、车侧PWM信号注入（充电桩侧身份欺骗，影响充电电流大小）、潜在地CAN总线威胁。
 
 % 因为车辆的检测逻辑是优先检查插枪状态，攻击者可以实现远程控制的拒绝服务攻击。

% 攻击者通过物理接触到充电枪以及将车枪插入到行为是不需要车主预先授权的，也就是任何人，包括攻击者，都可以完成将充电枪插入电车充电口的行为。其中涉及到攻击者接触到充电枪（公共充电桩），充电口（在锁车状态下一旦接受到充电枪的无线信号充电口就会自动打开），对车辆也不需要是启动或者刚关不久的状态，也就是插枪动作不需要车主授权，任何人都可以在任何时刻将充电枪插进锁住车门的电车当中。

\bsub{Roadmap.}
 This paper is organized as follows: \S~\ref{Sec:Preliminary} presents the fundamental concepts of the EV charging process, establishing the technical groundwork for understanding vulnerabilities in the system. \S~\ref{Sec:overview} explores the weak authentication vulnerabilities associated with the CC \& CP ports in EV Charging system.
 In \S~\ref{Sec:Portulator}, we introduce \ourwork, a novel device that can be discreetly installed on different standards of charging guns, enabling various real-world attack scenarios. \S~\ref{Sec:Implementation} provides an in-depth demonstration of three practical attacks to validate the effectiveness of \ourwork. \S~\ref{Sec:related} reviews relevant research on signal spoofing and its impact on charging port security. In \S~\ref{Sec:countermeasures}, we propose two defense mechanisms against CC port impedance manipulation and PWM signal spoofing. \S~\ref{Sec:discussion} discusses the ethical considerations, limitations of \ourwork, and potential directions for future research.  Finally, \S~\ref{Sec:conclusion} summarizes the key findings regarding weak authentication vulnerabilities in EV charging systems.

% \section{Background}  
\section{Preliminaries} \label{Sec:Preliminary}

The advent of electric vehicles (EVs) as a sustainable alternative to traditional fossil fuel-powered automobiles has necessitated the development of a robust and efficient charging infrastructure. This infrastructure is supported by various charging port technologies and authentication protocols designed to facilitate EVs' safe and effective charging. The diversity in charging standards, including GB/T 20234(predominantly used in China), IEC EU (International Electrotechnical Commission) standards for European compatibility, SAE J1772 (common in the United States and other countries), NACS (North American Charging Standard) and the Combined Charging System (CCS) catering to both AC and DC charging, reflects the global effort to enhance EV accessibility and utility. Each standard specifies physical guns and electrical specifications, ensuring compatibility and safety across vehicles and charging piles.

\subsection{Security Risks in EV Charging System}

The EV charging process involves several key steps, beginning with the removal of the charging gun, which activates a wireless sensor and triggers the vehicle's charging port lid to open, as discussed in \S~\ref{Section:sensor_back}. 
\vspace{-8px}

\begin{figure}[h]
    \centering
    \includegraphics[width=0.8\columnwidth]{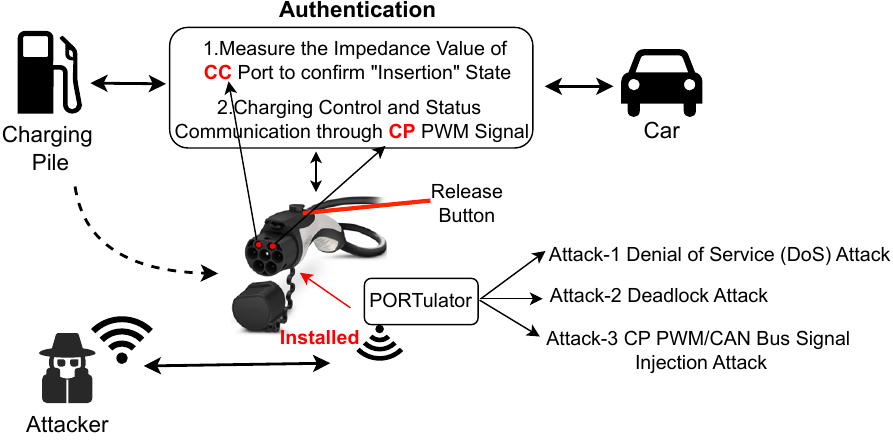}
    \caption{Overview of \ourwork Attack Vectors on EV Charging Infrastructure}
    \label{fig:Button}
    \vspace{-10px}
\end{figure}
\vspace{-5px}
    
Once the charging gun is connected to the vehicle’s charging port, a series of communications is initiated between the vehicle and the charging pile. The vehicle checks the \textbf{Charging Confirmation (CC)} port, \textbf{Proximity Pilot (PP)} port, or in the case of fast charging, the CC1 and CC2 ports, to verify the connection status of the charging gun. Once these ports detect a valid connection (i.e., the ports are no longer in an "open" state), the vehicle’s charging port engages a locking mechanism that secures the charging gun in place. This lock, an electromechanical device, ensures that the charging gun remains firmly attached during the charging process. The details of this locking mechanism are further discussed in \S~\ref{Section:safety}.

Furthermore, the vehicle transmits essential information, such as the Vehicle Identification Number (VIN) and frame number, to the charging pile through the Control Pilot (CP) port. This communication is particularly common with certain manufacturers. More critically, the CP port facilitates the exchange of the vehicle’s status, required charging power, and current specifications to regulate the charging process.

Throughout the charging process, the vehicle and charging pile maintain continuous communication via the charging cable to ensure the charging process runs smoothly. Once the battery reaches full charge, the vehicle notifies the charging pile via the CP port to end the session. In public charging stations, after payment is processed, the user can disconnect the charging gun by pressing the release button on the gun, which deactivates the locking mechanism and allows the gun to be safely removed, as shown in Figure~\ref{fig:Button}. The user then returns the charging gun to the pile and is free to leave.

% 充电枪的认证：
%   充电枪插入到充电口上时， 通常用户会释放枪体上的按键这个结构，这个结构不单单是一个物理的卡锁装置，这个物理按钮的press and release 意味着内部电路的改变。 具体而言，主要改变CC 线路的线信号，充电枪被插在充电口处之后

\subsection{Weaknesses in Chargers' Authentication Process}

\begin{figure*}[!ht]
\setlength{\abovecaptionskip}{0.cm}% 0.0cm}
    \centering
    \subfloat[GB/T AC \\ GB/T 20234.2]{
        \includegraphics[width=0.15\linewidth]{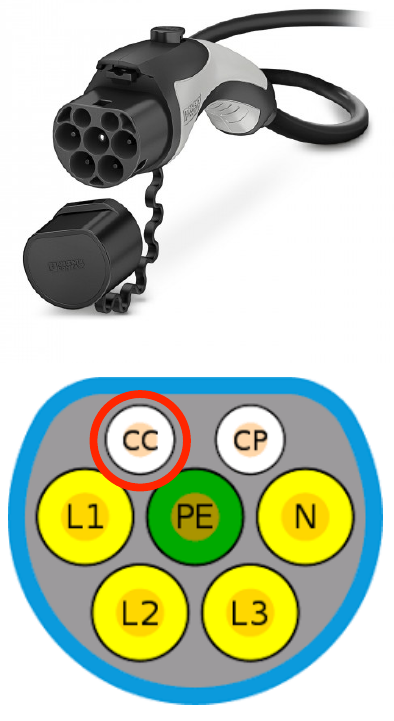}
        \label{fig:Charge_pinout:a}
    }
    \quad
    \subfloat[GB/T DC \\ GB/T 20234.3]{
        \includegraphics[width=0.16\linewidth]{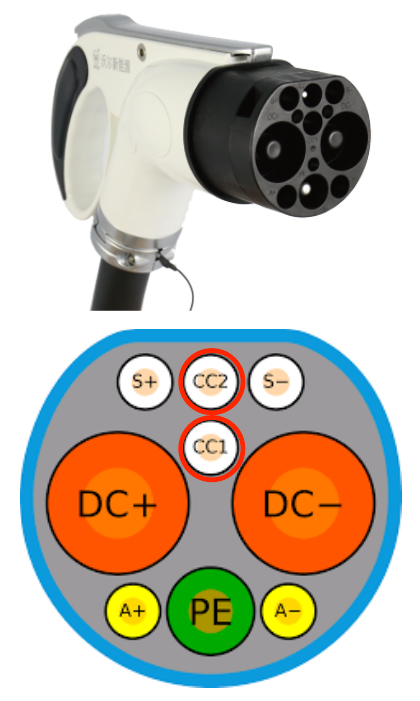}
        \label{fig:Charge_pinout:b}
    }
    \quad
    \subfloat[NACS]{
        \includegraphics[width=0.2\linewidth]{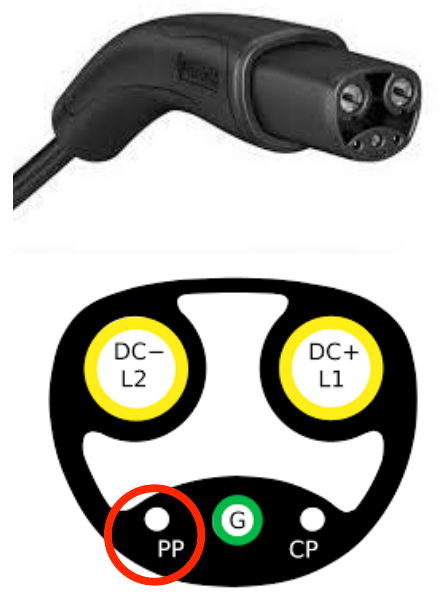}
        \label{fig:Charge_pinout:c}
    }%\vskip -10pt
     \quad
    \subfloat[SAE J1772]{
        \includegraphics[width=0.17\linewidth]{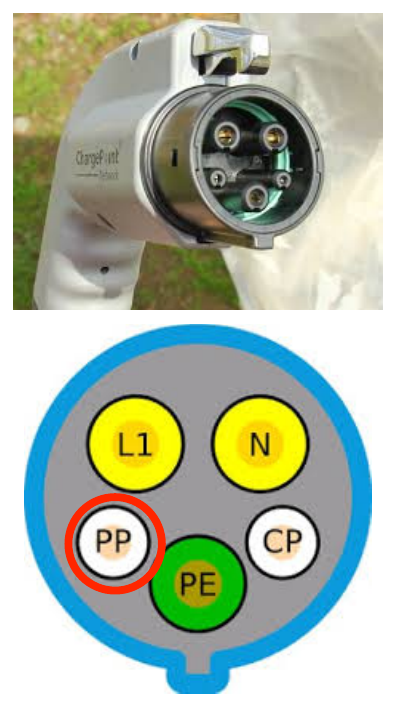}
        \label{fig:Charge_pinout:d}
    }
    \quad
    \subfloat[CCS I]{
        \includegraphics[width=0.13\linewidth]{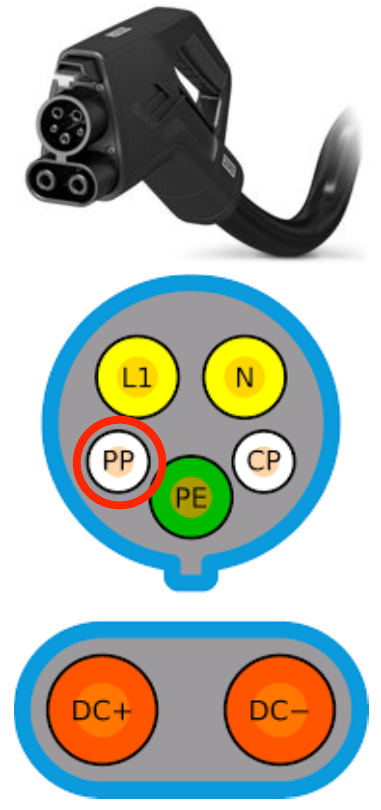}
        \label{fig:Charge_pinout:e}
    }
    \quad
    \subfloat[IEC 62196]{
        \includegraphics[width=0.17\linewidth]{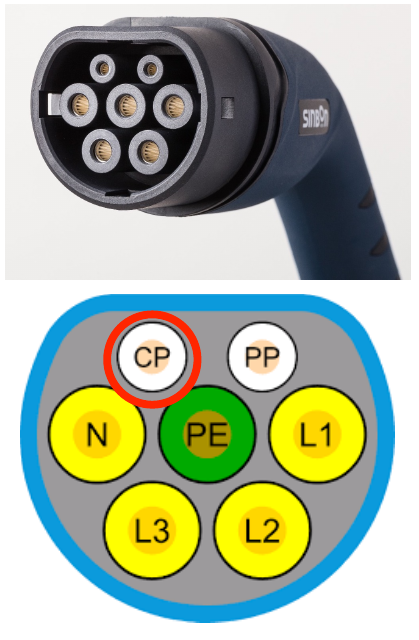}
        \label{fig:Charge_pinout:f}
    }
    \quad
    \subfloat[CCS II]{
        \includegraphics[width=0.35\linewidth]{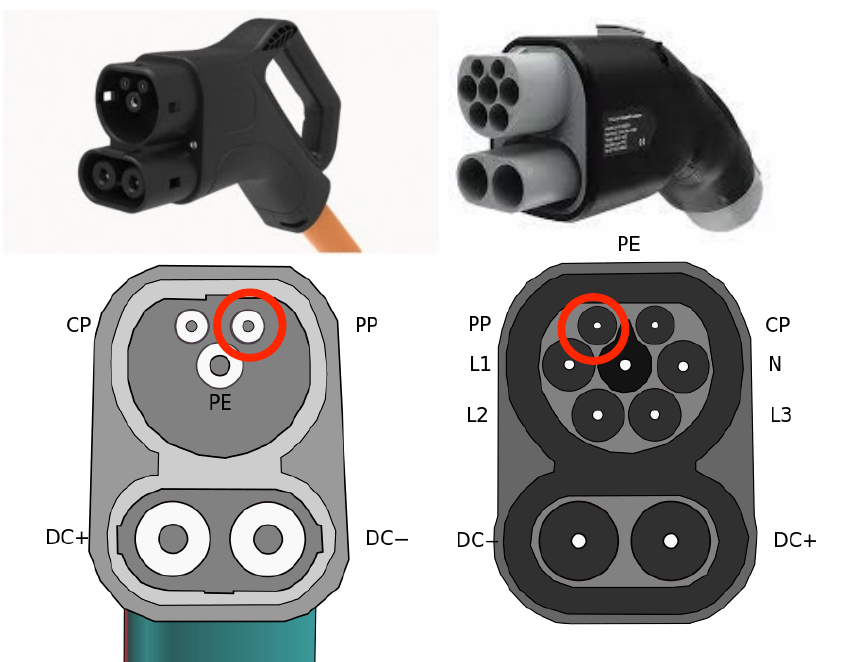}
        \label{fig:Charge_pinout:g}
    }
    \quad
    \caption{Charging Gun Standards (Signals from the ports, highlighted in red circles, confirm the proper connection between the EV and the charging pile.)}
    \label{fig:Charge_pinout}
    \vspace{-5px}
\end{figure*}

The authentication process begins when the charging gun is inserted into the vehicle’s charging port. Typically, the user presses and releases a button on the gun, which not only physically engages the lock but also triggers a signal change in the internal circuit. Specifically, this action alters the signal on the CC line, signaling that the gun has been properly inserted into the vehicle's port. As shown in Figure~\ref{fig:Charge_pinout} and Table~\ref{tab:resistor_command}, all charging standards share a similar connection confirmation step during the authentication process.

However, this mechanism introduces a critical weakness. By spoofing the expected resistance on the CC line (highlighted in red), an attacker can trigger a “deadlock” condition: the EV mechanically locks the charging gun before any digital authentication occurs. Once locked, further interaction is blocked, halting the process and creating a denial-of-service state. This flaw is present across standards, exposing a shared point of failure in physical-layer authentication.

Before electrical power is transferred, the system verifies the connection to ensure safety. A voltage detection point in the CC port circuit continuously monitors real-time changes to confirm the proper connection. Once the correct signal value is detected, the system advances to the next step: identification and the handshake protocol.

\vspace{-15px}

 \vspace{-7px}

Identification and handshake between the charging pile and the vehicle are facilitated by the CP port, through which data packets are exchanged. These packets verify the identities of both the car and the charging pile, ensuring compatibility and preventing any electrical mistakes. During the process, the vehicle communicates its electrical requirements, which are cross-checked with the charging pile’s capabilities. This prevents mismatches that could lead to unsafe conditions.
In cases where the vehicle is part of a charging network or requires user authentication, the process may involve further communication with a central charging management system. This ensures that the vehicle or user has the proper authorization to access the charging service. In some instances, this may include confirming payment authorization before charging begins.

Once authentication and authorization are complete, the vehicle transmits its specific charging needs to the pile, and the charging session begins. Continuous monitoring by both the vehicle and the pile ensures that electricity is delivered safely and efficiently, with adjustments made dynamically based on the vehicle's requirements and the charging pile's capabilities.
At the end of the charging session, a final communication between the vehicle and the charging pile safely terminates the process. The charging port's locks are then disengaged, allowing the user to safely remove the charging gun.

\subsection{Control Signals in Charging Pile} \label{Section:sensor_back}

 \bsub{Wireless Signal for Opening the Charging Port Lid.} Modern EVs often support wireless unlocking of the charging port lid to improve usability. This is commonly triggered when the user presses a button on the charging gun, which transmits a fixed wireless “open lid” signal to a receiver near the charging port.

\begin{figure}[!ht]
     \centering
     \includegraphics[width=0.8\linewidth]{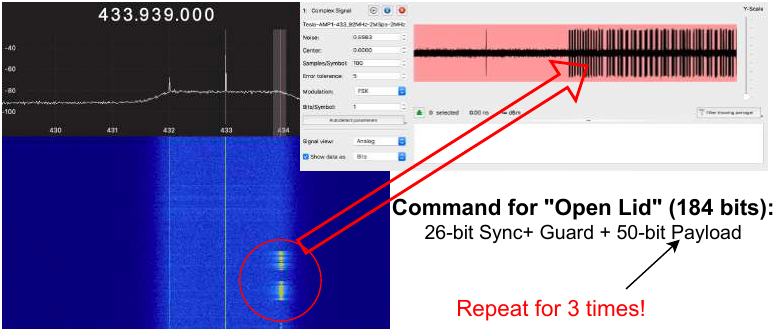}
     \caption{Wireless Signal for "Open Lid"}
     \label{fig:wireless}
     \vspace{-5px}
 \end{figure}

 As shown in Figure~\ref{fig:wireless}, we captured this signal from a NACS-standard charging gun using a HackRF One and GQRX. Each trigger emits ten identical packets, each containing a 26-bit sync word followed by three payloads, delimited by 3-bit guards. The last bit in the final packet is always `0`, indicating transmission end.

 \vspace{-5px}

 Using Universal Radio Hacker (URH), we replayed this signal on multiple EV models, including Tesla Model S/Y and Volkswagen ID.4. In every case, the charging port lid opened without needing any authentication. Notably, GB/T-compliant Tesla Superchargers in China use the exact same signal format.

 This uniformity reveals a critical flaw: the wireless “open lid” command is a fixed, unauthenticated signal reused across platforms. As a result, an attacker with low-cost equipment can remotely gain physical access to the charging port, enabling further attacks on the vehicle's internal interface.

\bsub{PWM Signal for Controlling Charging Current.}
\modify{A 1\,kHz Pulse Width Modulated (PWM) signal is transmitted through the CP line to coordinate charging power. The duty cycle of this signal is defined as $Duty Cycle = \frac{Time ON}{Total Period} X 100\%$. 
It determines the proportion of time the signal remains high during each cycle and directly reflects the available current capacity~\cite{shi2024laser}. For example, a 50\% duty cycle typically indicates an available current of 32\,A at 220–250\,V AC, corresponding to approximately 7–8\,kW of charging power.}

\modify{The vehicle interprets both the duty cycle and the peak voltage level of the CP signal to determine the operational state, such as \textit{standby}, \textit{charging}, \textit{ventilation required}, or \textit{error}~\cite{iec61851,iec309,evparkcharge}. Further details on the authentication and signal interpretation process are discussed in Section~\ref{Section:auth}.}

  % \begin{figure}[!ht]
  %     \centering
  %     \includegraphics[width=0.85\linewidth]{figure/CP_PWM.pdf}
  %     \caption{CP Signal Regulating Charging Current}
  %     \label{fig:CP_PWM}
  %     \vspace{-5px}
  % \end{figure}

  % These states are as follows: (1)State A (Peak voltage with 12V): Standby mode, indicating the vehicle is not connected. (2)State B (Peak voltage with 9V): The charging gun is connected, but charging has not yet begun. (3)State C (Peak voltage with 6V): Charging has started. (4)State D (Peak voltage with 3V) Charging is in progress, but ventilation is required for indoor areas. If the signal’s peak voltage deviates from the expected values (e.g., the valley does not reach -12V), the system detects an error state~\cite{iec61851,iec309,evparkcharge}.
   % The charging pile determines the charging current based on the duty cycle of the PWM signal, as transmitted by the vehicle. For instance, a 50\% duty cycle corresponds to a charging current of 32A. Given an AC voltage range of 220–250V, this translates to a charging power of approximately 7–8kW. 

\subsection{Safety Measures for EV Charging} \label{Section:safety}

Ensuring charging safety is essential for both EVs and chargers due to the high operating voltages.
To address this issue, vendors have implemented several safety measures, including emergency switches, electromechanical locks, and temperature sensors. In this section, we focus on two key safety features found in modern charging systems: \textbf{electromechanical locks} and \textbf{temperature sensors}, both of which work together to ensure safe and uninterrupted charging.

  \bsub{Electromechanical Locks for Preventing Electrical Hazards.} Electromechanical locks are a fundamental safety measure implemented across various charging standards, including GB/T, IEC, SAE J1772, NACS, and CCS~\cite{dcchargingprocess,TeslaModelXManual2024,wiki-ccs,wiki:SAE_J1772}. These locks provide a physical layer of security by preventing the charging gun from being accidentally disconnected during the charging process. This is especially important given that the voltage used in EV charging systems far exceeds the typical safety threshold for humans and animals, which is around 36V.

  \begin{figure}[!ht]
    \centering
    \includegraphics[width=0.8\linewidth]{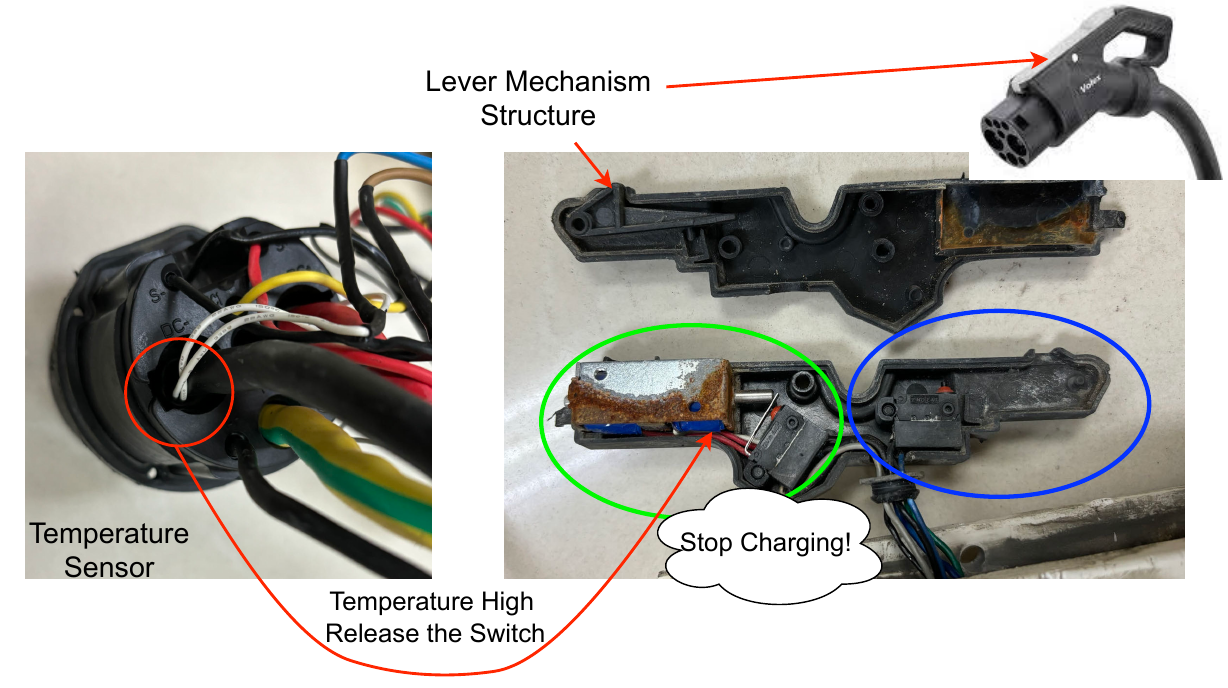}
    \caption{Automatic Triggering of High-Temperature Protection Switch}
    \label{fig:gun_structure}
    \vspace{-10px}
\end{figure}
  
  Once the charging gun is connected and the vehicle detects the correct CC port voltage, the lock engages, ensuring that the charging gun remains securely in place throughout the session. This prevents accidental disconnections and potential exposure to high voltage. In addition to safeguarding users from electrical hazards, the lock ensures continuous power transfer, which is essential for both safety and efficient charging.
  
  As shown in Figure~\ref{fig:gun_structure}, the travel switch in the blue circle monitors the connection status of the charging gun. When the "Unlock" button is pressed, the lever mechanism moves from a pressed to a raised state, causing the impedance between the CC1 and PE ports to drop to $0~\Omega$, signaling to the vehicle that the charging gun is ready to be removed. This ensures that the system correctly disengages the electronic lock and ends the charging session safely.

  \bsub{Temperature Sensors for High-Temperature Protection.} In addition to electromechanical locks, temperature sensors play another vital role in ensuring the safe operation of EV charging systems. According to standards such as IEC 62196 and GB/T 20234.3 for DC fast charging,  the internal temperature of the charging system must not exceed 90°C. Overheating can lead to significant safety risks, including damage to both the vehicle battery and the charging equipment, making temperature monitoring a key safety feature.

  As shown in Figure~\ref{fig:gun_structure}, temperature sensors (highlighted in red) are strategically placed inside the charging gun to continuously monitor heat levels. If the temperature exceeds the safety threshold, these sensors trigger an automatic response. The thermal protection mechanism (marked in green) activates the motor, causing the left travel switch to close. This sends a stop signal to the charging pile, halting the charging process to prevent further overheating.

  \vspace{-10px}

\section{Overview} \label{Sec:overview}

\subsection{Motivation}\label{Section:motivation}
% 我们的motivation源于GB/T AC充电枪的漏洞是否在其他充电接口标准上也存在的疑问。若这是一种在世界上各类型充电枪上普遍存在的漏洞则会有严重的安全威胁，意味着攻击者基于不同类型、不同状态所需的特征电阻值，便可以跨品牌、跨地区威胁电动车的充电体系。进一步，如果这个漏洞可以被远程控制触发，则意味着攻击者可以远程死锁指定充电枪与车辆，实现如勒索赎金等恶意目的。

The global adoption of EVs has accelerated the deployment of diverse charging infrastructures. 
Meanwhile, this rapid expansion introduces new security risks, especially in the physical and signaling layers of EV–charger communication~\cite{kohler2022brokenwire}.
Recent research has identified a weak authentication vulnerability in the GB/T 20234.2 standard~\cite{song2023ransom}, which enables adversaries to manipulate resistance values in the charging confirmation (CC) port, thus inducing unauthorized state transitions or even deadlocks between vehicles and charging piles.

\modify{ However, the potential impact of such attacks remains unclear, especially given the diversity of charging standards. 
A critical question arises: Can malicious signals be injected across all charging standards, and if so, what are the broader consequences? 
Beyond simply disrupting charging, could such attacks target an EV's charging management system or even inject malicious commands into the EV's internal data bus? To address these concerns, we conduct an empirical study to investigate whether more devices with diverse charging standards are vulnerable to physical signal injection attacks and explore whether these attacks could result in more severe consequences.}

% \modify{
% Given the 
% This raises an urgent question: Are similar vulnerabilities present across other international charging standards? If so, attackers could exploit differences in resistor values and state logic to launch cross-brand, cross-standard denial-of-service or ransom-style attacks. Moreover, if such attacks can be triggered remotely, they pose a serious threat to the public charging infrastructure.}
% Motivated by this possibility, we investigate whether such signal-level weaknesses are systemic across protocols, and explore the feasibility of remote exploitation through spoofed impedance and PWM/CAN bus signal injection at the charging interface.

%\vspace{-10px}

\subsection{Weak Authentication Vulnerabilities in Charging Port} \label{Section:auth}
% 可能主要讲述充电强内部电路逆向的方法，检测端口之间的电气隔离：
% 使用绝缘电阻测试仪或万用表的连续性测试功能。
% 测量此端口与设备其他端口之间的电阻。
% 如果读数显示高阻抗或无限，说明端口之间电气上是隔离的。

\begin{figure}[!ht]
    \centering
    \includegraphics[width=0.9\linewidth]{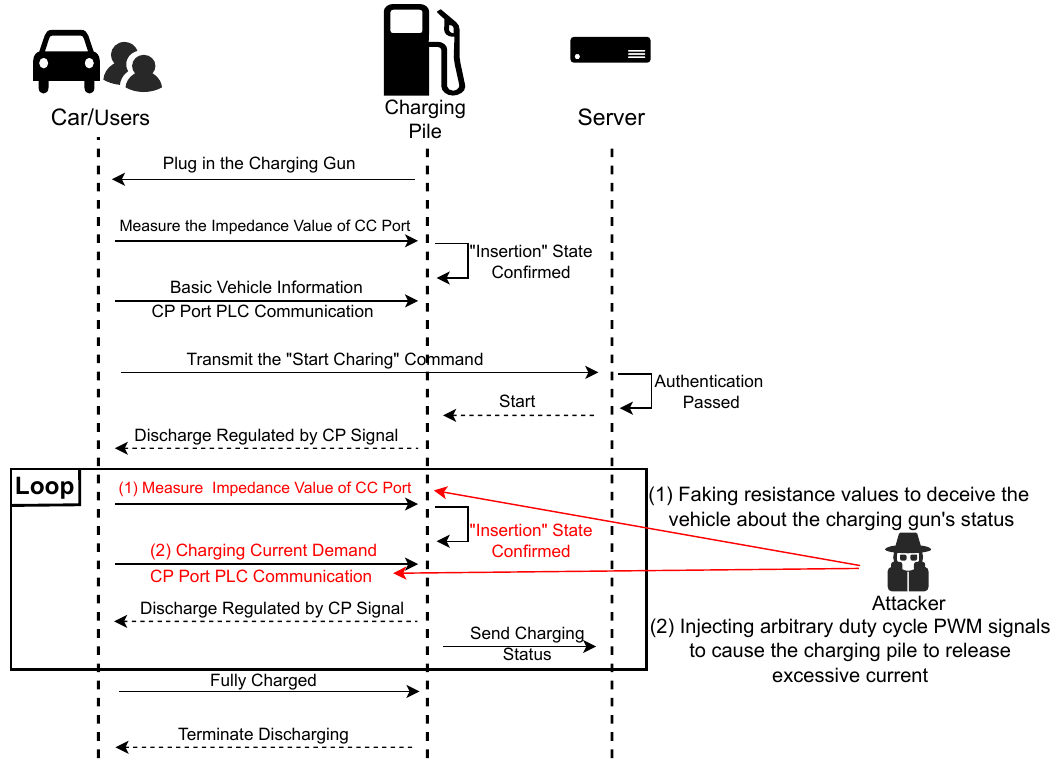}
    \caption{Falsified Signal Attack Exploiting Weak Authentication in Charging Protocols}
    \label{fig:protocol}
    \vspace{-5px}
\end{figure}

\modify{Authentication mechanisms in EV charging protocols are designed to prevent unauthorized access to vehicle charging functions. However, our investigation reveals that several widely adopted standards rely on weak signal level authentication~\cite{HackerNews2023}, where charging state transitions are determined by analog parameters such as port impedance or PWM duty cycles, without any cryptographic validation. This architectural assumption leaves room for adversaries to spoof authentication signals and gain unauthorized control over the charging process.} Figure~\ref{fig:protocol} illustrates a typical falsified signal attack that exploits this weakness.

To explore this vulnerability, we begin by reverse engineering the internal signal circuits of several representative charging guns. In an initial study of a GB/T AC charging gun, we use a multimeter to probe the impedance of the CC line and identify an unexpectedly simple configuration consisting of only two resistors and a mechanical travel switch. Although this demonstrates the feasibility of spoofing CC signals by manipulating resistor values, we also note the risk of signal coupling between ports. To better understand the design, we disassemble the gun and confirm that the authentication logic relied purely on passive resistance and mechanical actuation, with no built-in tamper detection.

Since many commercial charging guns employ anti-tamper designs that hinder physical disassembly, we developed \ourwork, a non-invasive port analysis tool capable of automatically identifying internal electrical characteristics. By interfacing with multiple charging gun ports, \ourwork can assess port isolation, impedance values, and the presence of memory components such as capacitors or inductors, without opening the casing. This allows us to extract authentication logic from devices across multiple standards in a repeatable and scalable manner.

\begin{figure}[!ht]
    \centering
    \subfloat[Slow Charging Gun Authentication Circuit]{
        \includegraphics[width=0.45\linewidth]{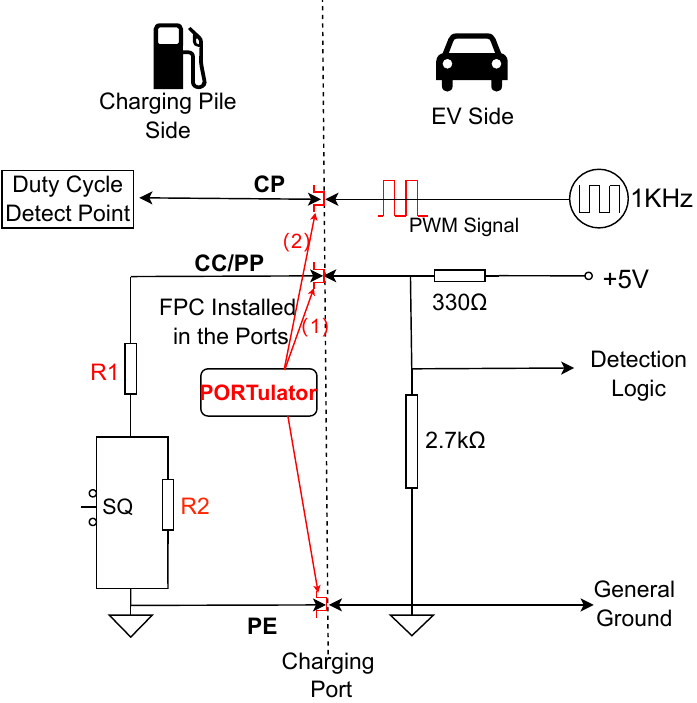}
        \label{fig:Inner_AC}
    }
    \hspace{8pt}
    \subfloat[Fast Charging Gun Authentication Circuit]{
        \includegraphics[width=0.47\linewidth]{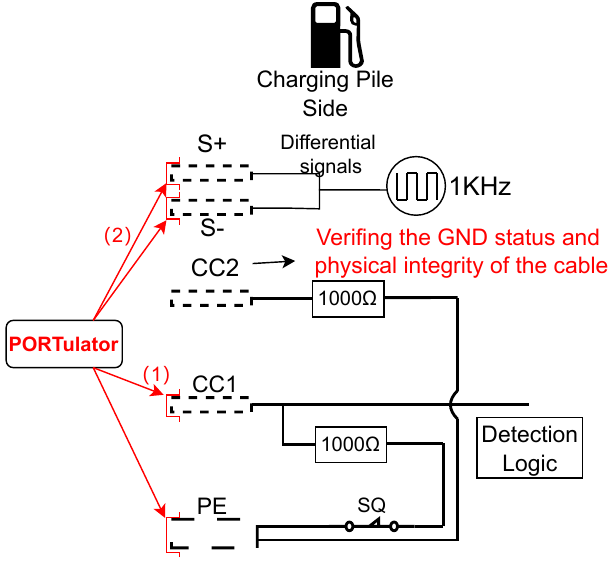}
        \label{fig:Inner_DC}
    }
    \caption{Comparison of Slow and Fast Charging Gun Authentication Circuits}
    \label{fig:Inner}
\end{figure}

Figure~\ref{fig:Inner} shows the typical internal signal circuits for slow and fast charging guns, which were mapped using our automated method. These diagrams reveal that signal-based authentication remains consistent across protocols: electrical states are inferred through analog conditions, rather than protocol-level cryptographic handshakes.

Building on this insight, we designed a signal spoofing attack targeting these weak authentication mechanisms. The attack involves mimicking the electrical signatures that represent various charging states. For example, as summarized in Table~\ref{tab:resistor_command}, changing the resistance value on the CC line can simulate events such as plug insertion, button press, or user confirmation. By replicating these states, an attacker can deceive the charging pile into initiating or continuing a charging session without authorization.

To construct these spoofed signals, we first used \ourwork to capture reference waveforms and parameter ranges during normal charging sessions. This included measuring resistance transitions, PWM frequencies, and voltage thresholds under different operational states. Based on this data, we generated counterfeit signals using a programmable MCU that emulates the behavior of a legitimate EV-side interface. These signals are then injected into the CC or CP ports at specific phases of the charging handshake to trigger unauthorized transitions.

Since the charging pile relies solely on analog signal conditions for authentication, it is unable to distinguish between genuine and spoofed interactions. As a result, the attacker can initiate, manipulate, or deadlock the charging process, even without any access to cryptographic credentials or prior pairing with the vehicle.

\subsection{Threat Model}
% 强调是短时间接触设备即可。 然后对于设备隐蔽性的刻画放在section 4 设计这边， 枚举了同样接触改装形式攻击。

\modify{We consider a realistic adversary targeting public EV charging infrastructure, which is widely deployed in semi-supervised environments such as parking lots, apartment complexes, and service stations. These locations often lack strict physical supervision, allowing attackers \textbf{short-term access} to charging equipment. This model reflects real-world scenarios observed in similar physical-layer attacks, such as ATM skimming~\cite{ATM1,ATM2} and RFID spoofing~\cite{RFID}, where covert hardware can be deployed without attracting attention.} %\todo{R.4 Revised threat model to emphasize \textbf{short-term physical access} and realistic attack settings.}

The attacker is assumed to have brief physical access to a charging gun, either during their own charging session. Leveraging this opportunity, the attacker discreetly installs a modified version of \ourwork, which embeds signal injection hardware into the head or sheath of the charging gun. The device remains dormant until remotely triggered via wireless communication.

Once a victim connects their vehicle to the compromised charging gun, the attacker activates the device to inject falsified CC or CP signals. These spoofed signals exploit weak authentication mechanisms to induce denial-of-service conditions, lock the charging port, or manipulate current flow. \modify{In standards such as GB/T 20234.3 and NACS, the attack may further escalate to in-vehicle CAN Bus injection via exposed communication lines.}

This threat model does not assume firmware modification or charger disassembly. Instead, it demonstrates that brief, opportunistic access-combined with a camouflaged device, can enable powerful attacks in realistic public settings.

\section{\ourwork Design} \label{Sec:Portulator}

\modify{
To verify if the chargers are vulnerable to physical signal injection attacks, we propose \ourwork, a customize hardware platform based on the RP2040 Microcontroller Unit (MCU)~\cite{raspberrypi2023rp2040}, designed to uncover and exploit signal-level vulnerabilities in EV charging infrastructures. This device enables remote and precise manipulation of physical-layer communication between electric vehicles and charging piles, supporting real-world spoofing and injection attacks.}

\subsection{Hardware Design} 

\begin{figure}[!ht]
    \centering
    \includegraphics[width=0.95\linewidth]{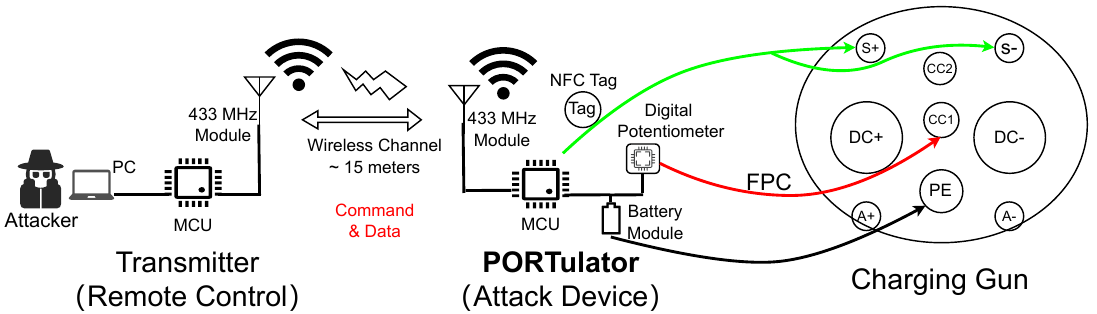}
    \caption{Design of \ourwork for Resistor Spoofing \& Signal Injection}
    \label{fig:portulator_design}
    \vspace{-5px}
\end{figure}

\modify{The core of \ourwork is a compact, modular spoofing device—\ourwork—built to physically interface with the CC and CP lines of standard EV charging guns. As shown in Figure~\ref{fig:portulator_design}, the system is powered by an RP2040 MCU, chosen for its real-time control capabilities, low-latency GPIO access, and flexible ADC/DAC integration.}

\modify{The PCB is designed to interface directly with both analog signaling pins (CC/CP) and digital monitoring subsystems. A programmable potentiometer (AD5160 module) is included to emulate impedance-based logic states on the CC line, while a PWM-capable GPIO output pin synthesizes the CP signal to reflect various charging states.
To allow for remote-controlled behavior, the hardware integrates a 433\.MHz wireless receiver (GC433-TC007) that accepts over-the-air commands from an Arduino-based controller. This setup enables dynamic payload delivery, such as adjusting resistance values or toggling CP duty cycles, effectively changing the perceived EV state in real-time.}

\modify{The physical device is encapsulated in a modified charging gun shell. Specifically, a thin custom cable is routed through the charging gun to the CC pin, internally connected to a pre-configured resistor, and routed through an insulating sleeve to avoid interfering with normal charging pile operations. A small metal ring is used to stabilize the CC contact position. This modification is minimally intrusive, does not affect standard charging under normal conditions, and is nearly invisible from the outside, making the attack device covert and practical for deployment in semi-public scenarios. In addition, the compact design enables rapid installation: the entire module can be integrated into a fake adapter or portable testing tool and clipped onto the target charging gun in under 90 seconds, minimizing the attacker's exposure time on-site.}

\begin{figure}[!ht]
    \centering
    \includegraphics[width=0.5\linewidth]{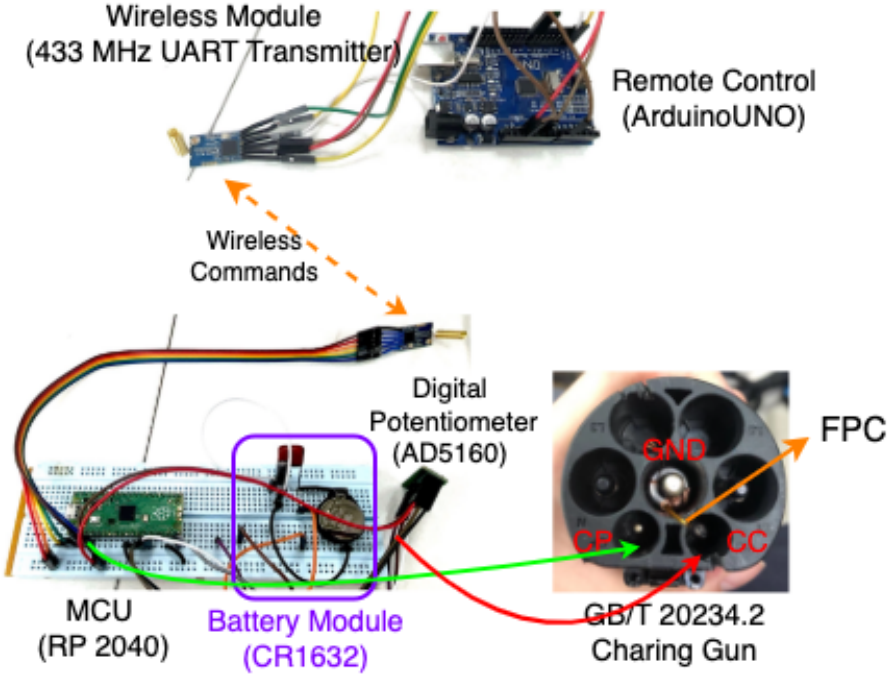}
    \caption{Physical Prototype of the \ourwork Attack Device}
    \label{fig:Detailed_device}
    \vspace{-5px}
\end{figure}

\modify{Figure~\ref{fig:Detailed_device} shows the physical construction of the prototype. All components used are off-the-shelf and reproducible: the microcontroller board (RP2040), AD5160 potentiometer, GC433 wireless module, and the modified charging gun enclosure. \ourwork 's hardware setup and open-sourced code is provided at: \url{https://github.com/Vehicle-Security}.} %\todo{R2. Expanded hardware design description of PORTulator to clarify component usage, charging gun modification, and deadlock interaction.}}

\vspace{-5px}

\subsection{Signal Interpretation \& Injection Principles} % 原理 + 图解

\begin{figure}[!ht]
    \centering
    \subfloat[Impedance-Based Logic for CC Port Status]{
        \includegraphics[width=0.46\linewidth]{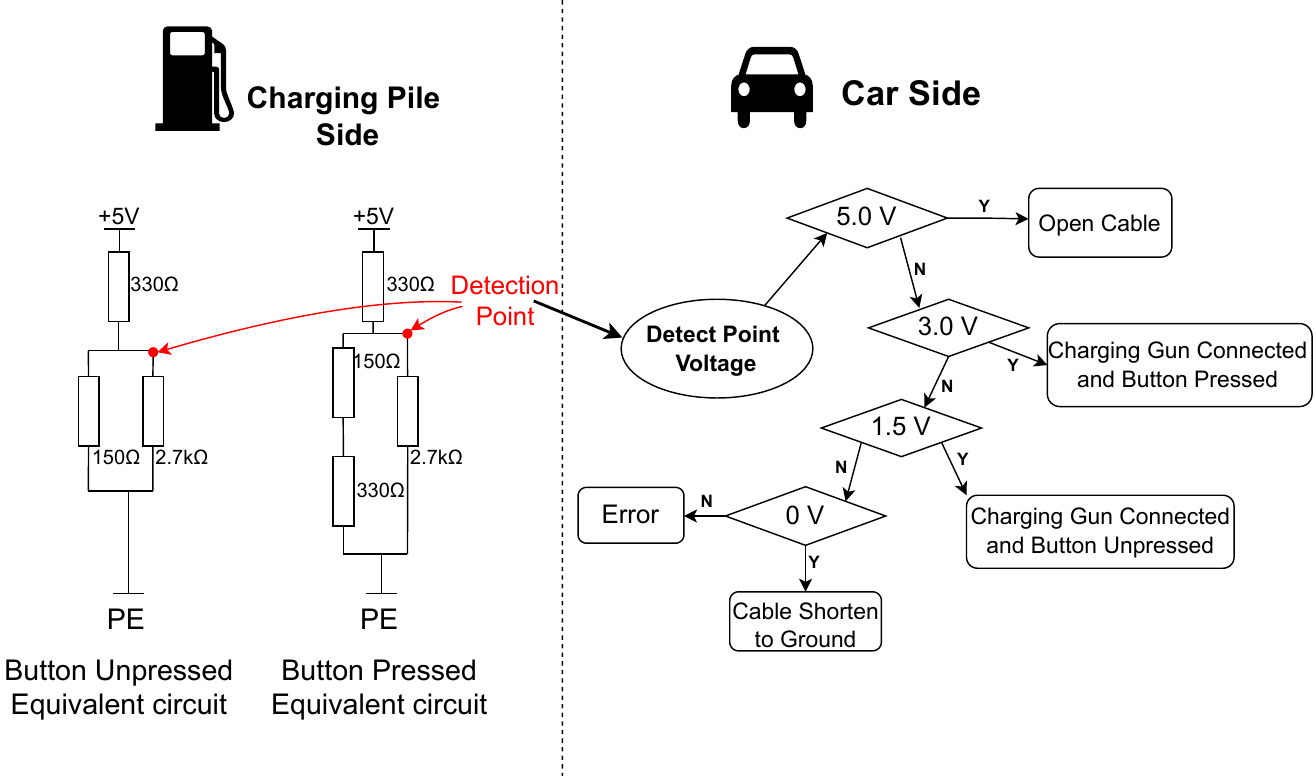}
        \label{fig:CC_impedance}
    }
    \hspace{5px}
    \subfloat[PWM Signal-Based Logic for CP Port Status]{
        \includegraphics[width=0.46\linewidth]{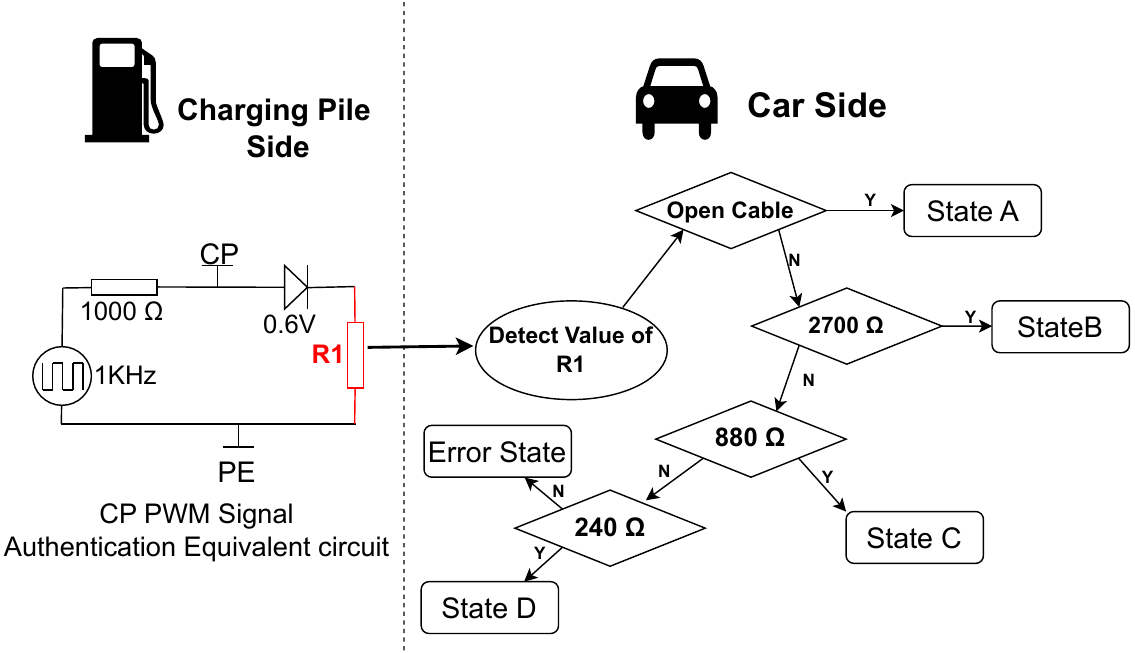}
        \label{fig:CP_logic}
    }
    \caption{\centering Parameter Values and Logical State Determinations at the CC Port (a) and CP Port (b) of EV Charging Gun}
    \label{fig:Portulator_test}
\end{figure}

\modify{The design of \ourwork is grounded in the signal-level understanding of EV charging protocols, particularly the logic interpretation on CC and CP lines. As illustrated in Figure~\ref{fig:Portulator_test}, our system mimics legitimate interactions by matching impedance and PWM behaviors expected during the communication phase.}

\vspace{-5px}

\begin{itemize}

\item \textbf{Impedance-Based Logic for CC Port Status.}
\modify{The CC pin voltage is determined by a resistive voltage divider formed between the EV's internal pull-down resistor and a fixed resistor inside the charging gun, often connected through a travel switch linked to the physical button. As shown in Figure~\ref{fig:CC_impedance}, the EV measures the voltage between the detection point (CC port) and GND to infer the connection state: 5\,V indicates an open cable, ~3\,V means the charging gun is connected and the button is pressed, ~1.5\,V indicates connected but unpressed, and 0\,V signals a fault or short condition.}

\item \textbf{PWM Signal-Based Logic for CP Port Status.}
\modify{The CP line enables the charger to determine the vehicle’s connection and charging states based on the equivalent resistance $R_1$ presented between the transistor and ground, as shown in Figure~\ref{fig:CP_logic}. In the default unconnected state (State A), no resistor is applied, and the charger interprets this as “cable not connected.” Upon physical connection (State B), the vehicle applies a $2.74k\Omega$ pull-down resistor to indicate the presence of the EV without charging intent. When the vehicle is ready to charge (State C), it adds a $1.3k\Omega$ resistor in parallel, forming an equivalent resistance of approximately $880\Omega$. This resistance signals to the charger that charging is authorized.
Only in State C does the vehicle actively respond by modulating the CP line with a 1\.kHz PWM signal. The duty cycle of this signal encodes the maximum allowable charging current (e.g., 50\% for 32A, 85\% for 51A). Additionally, certain implementations apply a $240\Omega$ equivalent resistance to indicate DC charging with forced ventilation requirements.
Any deviation from the expected resistance values, such as short or undefined configurations, results in a fault condition (State D). Overall, the EV communicates its charging state not by direct signaling, but by dynamically altering the resistance on the CP line.}

\end{itemize}

To ensure compatibility across various charging standards, we systematically reproduced the expected impedance values used to signal connection states. As shown in Table~\ref{tab:resistor_command}, \ourwork precisely emulates these reference resistances, with minimal deviation, to maximize spoofing success across a range of charging standards. This calibration enhances cross-standard reliability and enables consistent behavior in both AC and DC charging scenarios.

\begin{table}[!ht]
\centering
\resizebox{0.95\textwidth}{!}{%
\begin{tabular}{@{}ccccc@{}}
\toprule
\textbf{Standard} & \textbf{\begin{tabular}[c]{@{}c@{}}Unpressed Status\\ (Expected Impedance)\end{tabular}} & \textbf{\begin{tabular}[c]{@{}c@{}}Real Impedance \\  (Deviation $\Omega$)\end{tabular}} & \textbf{\begin{tabular}[c]{@{}c@{}}Pressed Status\\ (Expected Impedance)\end{tabular}} & \textbf{\begin{tabular}[c]{@{}c@{}}Real Impedance \\  (Deviation $\Omega$)\end{tabular}} \\ \midrule
SAE J1772     & $480~\Omega$                                                                             & 487 (+1.5\%)                                                                            & $150~\Omega$                                                                           & 145 (-3.3\%)                                                                            \\
\rowcolor[HTML]{EFEFEF} 
CCS I         & $480~\Omega$                                                                             & 487 (+1.5\%)                                                                            & $150~\Omega$                                                                           & 145 (-3.3\%)                                                                            \\
IEC 61851     & $1030~\Omega$                                                                            & 1027 (-0.3\%)                                                                           & $760~\Omega$                                                                           & 768 (+1.1\%)                                                                            \\
\rowcolor[HTML]{EFEFEF} 
CCS II        & $1030~\Omega$                                                                            & 1027 (-0.3\%)                                                                           & $760~\Omega$                                                                           & 768 (+1.1\%)                                                                            \\
NACS          & $460~\Omega$                                                                             & 466 (+1.3\%)                                                                            & $400~\Omega$                                                                           & 390 (-2.5\%)                                                                            \\
\rowcolor[HTML]{EFEFEF} 
GB/T 20234.2  & $220~\Omega$                                                                             & 210 (-4.5\%)                                                                            & $3520~\Omega$                                                                          & 3511 (-0.3\%)                                                                           \\
GB/T 20234.3  & $0~\Omega$                                                                               & 0 (0\%)                                                                                 & $1000~\Omega$                                                                          & 1003 (+0.3\%)                                                                           \\ \bottomrule
\end{tabular}%
}
\caption{\centering Comparison of Expected and Spoofed Impedance Values Across Different Charging Standards}
\label{tab:resistor_command}
\end{table}

\subsection{Adaptive Control \& Attack Interface} 

The adaptive control and attack interface of \ourwork is designed to facilitate real-time manipulation of spoofed signals during an ongoing attack. This interface integrates both low-level hardware control and a high-level user interface to enable precise and efficient execution of various spoofing scenarios.

At the hardware level, the MCU is responsible for processing incoming commands and adjusting signal outputs accordingly. These commands, typically sent in the form of HEX codes over a 433MHz wireless link, instruct the MCU to modify key signal parameters such as resistance on the CC line or duty cycle on the CP line. For instance, issuing a command to set the CC line resistance to a specific value allows the device to mimic the “charging plug inserted” or “charging authorized” states, effectively manipulating the EV-side logic as perceived by the charging pile. In addition to active control, the system supports real-time monitoring of the charging pile’s behavior during the attack. Embedded sensors measure key electrical parameters such as voltage and current, while communication feedback channels track the pile’s response to spoofed signals.

%To support flexible attack orchestration, a cross-platform graphical user interface (GUI) has been developed. This software interface allows users to configure signal parameters dynamically, select from a set of predefined spoofing modes, and store or recall configuration profiles for rapid deployment. The GUI is designed to be intuitive, providing real-time feedback on the signal values being transmitted and the state transitions being induced. This makes it feasible for attackers to manage complex interactions with minimal manual effort.

%When anomalies are detected, such as unexpected voltage dropouts or denial of handshake, the system can adapt its behavior automatically. Conditional logic routines within the MCU analyze incoming sensor data and adjust spoofed outputs in real time, maintaining the appearance of a legitimate EV even under changing conditions. This adaptive capability is crucial for sustaining the effectiveness of the attack throughout the charging session.

\section{Evaluation} \label{Sec:Implementation}
  %Implementation
  
 %充电桩欺骗电阻攻击
 %充电桩任意情况下的DoS攻击
 %理论分析的PWM信号注入情况分析

\modify{In this section, we evaluate the effectiveness of \ourwork across three key physical-layer attack scenarios: (1) inducing Denial-of-Service (DoS) conditions by manipulating the CC and CP lines, (2) spoofing resistor values to deadlock the charging gun in ransom-style attacks, and (3) injecting malicious PWM signals to manipulate charging behavior. We further explore the potential for higher-layer CAN Bus injection via the charging interface.}

\modify{Specifically, \ourwork successfully executed all three attacks across seven EV models and six major charging standards in Table~\ref{tab:experiment}, including GB/T 20234.2, GB/T 20234.3, CCS I, CCS II, NACS, and IEC 62196. These results highlight systemic weaknesses in EV charging infrastructures and underscore the need for stronger signal validation mechanisms across physical and communication layers.
Detailed information can be accessed at \url{https://github.com/Vehicle-Security}.}

\subsection{Case I: DoS Attack} \label{Sec:Case I}
%  车辆在充电过程当中会随时检测充电枪是否保持正常的插枪状态，这一点也十分合理，为了在高电压大电流的情况下充电桩放电不会造成车辆及人员的伤害，正如\ref{Sec:Case I}中 阐述的。根据\ref{Sec:Portulator}所提到PORTulator具备的功能，攻击者可以远程发送命令任意改变CC端口的电阻大小，也就意味着如果攻击者将端口阻抗改成异常电阻值，比如0欧姆或者代表着没插枪的大电阻数值，充电桩就会停止对EV充电。我们在大众ID.4, Telsa model S 都成功实现了中断充电的实验。 攻击者可以提前在公共充电桩上安装PORTulator，通过远程命令对多个充电枪标准完成拒绝服务攻击。

\begin{figure}[!ht]
    \centering
    \subfloat[DoS on CC to Disrupt Charging Process]{%
        \includegraphics[width=0.5\linewidth]{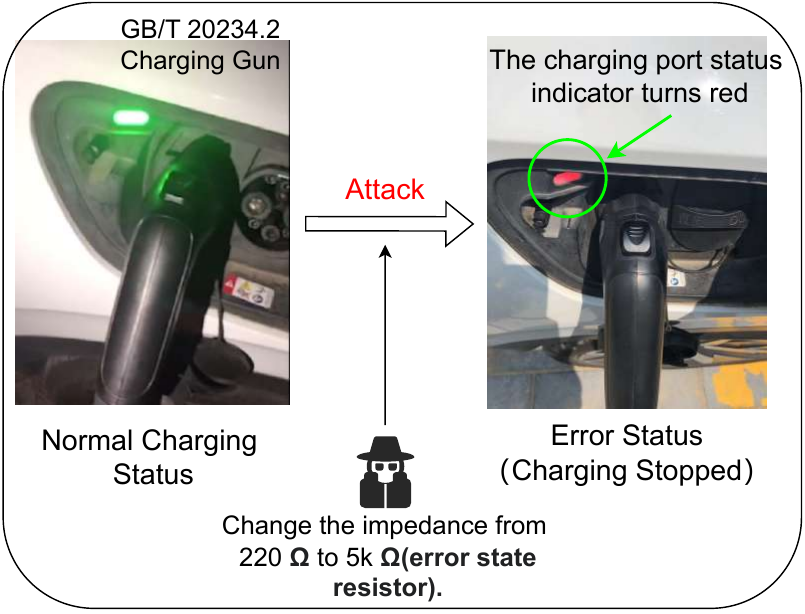}
        \label{fig:DoS_Attack_CC}
    }
    \hspace{5pt} 
    \subfloat[DoS on CP to Control Charging State]{%
        \includegraphics[width=0.41\linewidth]{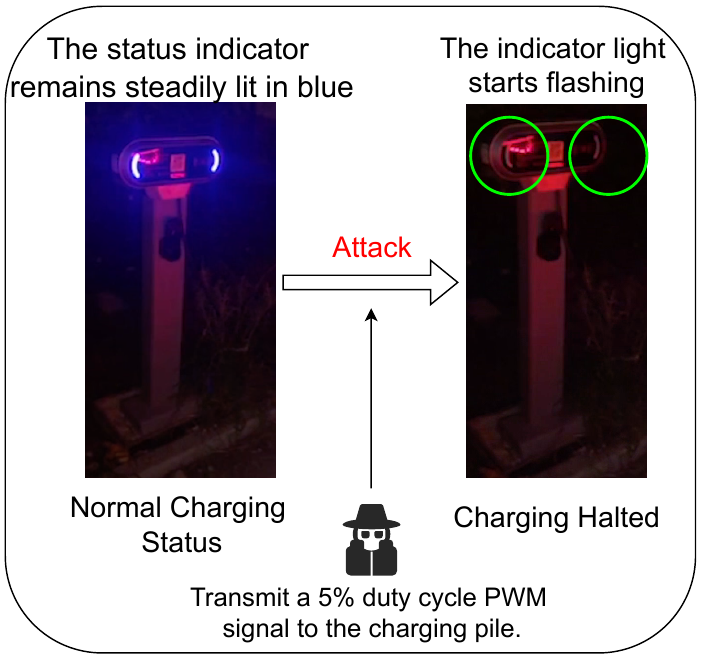}
        \label{fig:DoS_Attack_CP}
    }
    \caption{DoS Attacks on CC and CP Lines in EV Charging Systems}
    \label{fig:DoS_Attack}
    \vspace{-5px}
\end{figure}

EVs continuously monitor the physical state of the charging gun during charging to ensure safe operation under high-voltage and high-current conditions. Disruptions in the connection, such as improper insertion or unexpected impedance changes, are treated as critical events and will immediately halt the charging process.
As shown in Figure~\ref{fig:DoS_Attack_CC}, our attack leverages this safety mechanism by exploiting the capabilities of \ourwork, as detailed in \S~\ref{Sec:Portulator}. Specifically, attackers can remotely alter the CC line impedance to abnormal values, for instance, reducing it to 0 $\Omega$ or increasing it to simulate a disconnected state. This results in the charging pile immediately terminating the session. We validated this behavior on multiple vehicles, including the Volkswagen ID.4 and Tesla Model S. By pre-installing \ourwork on public charging piles, an attacker could remotely issue commands across various charging standards to launch broad DoS attacks.

Moreover, the CP line can also be targeted. As shown in Figure~\ref{fig:DoS_Attack_CP}, injecting a low-duty-cycle PWM signal, such as 5\%, can mislead the charging pile into interpreting the session as inactive communication or fault, further halting power delivery. This method offers another vector for reliably disrupting active charging sessions without requiring direct physical interaction.

\subsection{Case II: Deadlock Attack via CC Port Impedance Manipulation for Ransomware Exploitation} \label{Sec:Case II}

We further explore a novel form of ransomware attack that manipulates CC line impedance to trap the charging gun in a locked state, thereby immobilizing the EV and coercing users into making payments to regain control. Crucially, this attack does not rely on internet connectivity, distinguishing it from conventional ransomware campaigns~\cite{song2023ransom}.
% 一个优势、不同

\begin{figure}[!ht]
    \centering
    \subfloat[Deadlock Attack Applied to Ransom Attack]{%
        \includegraphics[width=0.51\columnwidth]{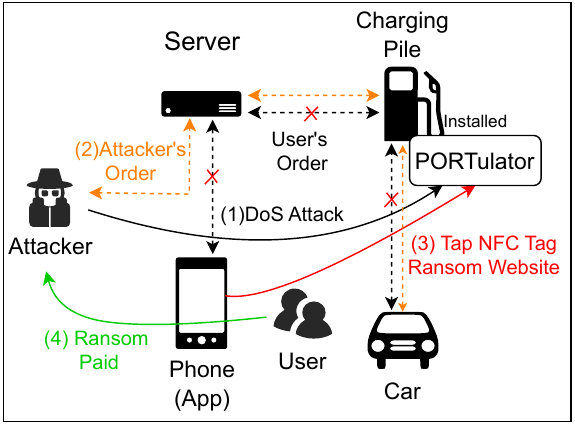} 
        \label{fig:ransom_process}
    }
    \hspace{5pt} % Adjust spacing between the two subfigures
    \subfloat[NFC Triggered Ransom Website Pop-up]{%
        \includegraphics[width=0.4\columnwidth]{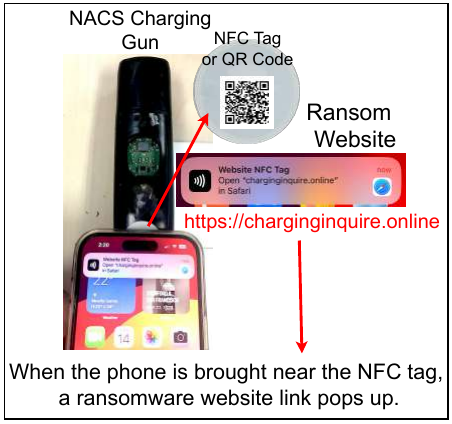} 
        \label{fig:NFC Triggered}
    }
    \caption{Deadlock Attack and NFC-Triggered Ransom Scenario}
    \label{fig:ransom}
    \vspace{-5px}
\end{figure}
\vspace{-5px}

As illustrated in Figure~\ref{fig:ransom_process}, the attack unfolds in four stages: 
(1)\textbf{DoS Attack}: 
The user initiates a legitimate charging session by connecting to a gun embedded with \ourwork. The attacker then remotely triggers a DoS condition (as outlined in \S~\ref{Sec:Case I}), halting the process by injecting abnormal signals on the CC or CP line.
(2)\textbf{Forged Charging Order Replay}:
Following disruption, the attacker replays captured CP signals from a previously observed session, initiating a new charging order that appears valid but is under attacker control (see \S~\ref{Sec:Case III}).
\modify{(3)\textbf{Ransom Prompt via NFC or QR Code}: 
When the user returns and attempts to stop the session or remove the charging gun, they find the interface unresponsive. Simultaneously, an NFC tag embedded in the charging gun, along with an optional QR code displayed on a sticker, triggers a prompt on the user's smartphone, as shown in Figure~\ref{fig:NFC Triggered}. This prompt opens a spoofed support webpage, such as "https://charginginquire.online", carefully designed to resemble a legitimate vendor site. Devices lacking NFC support can instead scan the QR code, which contains the same redirection metadata referencing NFC.}
(4)\textbf{Payment Demand}: 
The spoofed website informs the user that their charging session and thus their vehicle is locked. Instructions are provided to make a cryptocurrency or an anonymous payment to resume normal operation. This step effectively turns the charging infrastructure into a physical lever for extortion.

We successfully demonstrated this attack on a Volkswagen ID.4 using public charging piles operated by TELD and Starcharge in China. The exploit highlights a deeply concerning capability: attackers can gain remote control over EVs’ charging states, using infrastructure-side spoofing and interface deadlocks to extract payments from unsuspecting users. This underscores the urgent need for more robust authentication mechanisms, secure session management, and out-of-band validation to prevent such attacks. Detailed information and demo videos can be accessed at \url{https://github.com/Vehicle-Security}.

%这是一种基于攻击者构造特定状态的CC端口阻抗数值造成枪与车辆充电口死锁的效果所引发的一种新型勒索攻击，这种攻击相较于之前的工作ref{}是一种全新的不需要利用internet实现的攻击。Users begin their regular daily charging 选择了我们事先安装有\ourwork 的充电枪。攻击者（1）便可以通过case I中的任意一种方式，CC端口DOS攻击或者CP端口的DoS攻击打断当前的正常充电进程。（2）在打断原有的充电进程之后，攻击者便可以利用case III 中提到的Cp信号注入攻击，利用预先捕获的车辆与充电桩commuication的信号，重放这段信号完成开始充电前的basic information 交互。 之后便可以向充电桩注入特定占空的PWM信号开始充电，启动一个由attacker 控制的充电订单。（3）当用户返回准备停止充电，解锁离开时发现他们无权再控制当前的充电进程。我们提前在冲电枪上设计有一个内置了我们ransom website 网址的NFC Tag， 用户使用自己的手机tap 这个tag就会如Figure。。。 所示的一样弹出一个名字经过伪装的链接，for instance， https://charginginquire.online 伪装成厂商官方帮助使用者的咨询网站。 用户便会点击访问我们精心设计的勒索网页，在网页上用户将得知一条 Ransom message，当前车辆已经被锁定除非支付赎金否则便无法离开。（4）用户通过我们在ransom website上提供的链接，找到支付链接，完成赎金The attack involves manipulating server perceptions, deactivating charging orders, regaining control post-restoration, issuing ransom demands, and halting charging until payment. Experiments on Volkswagen ID.4 and public charging piles in China, including those by TELD and Starcharge, validate the method's effectiveness.

\subsection{Case III: CAN Bus Signal Injection Attack} \label{Sec:Case III}
% 修改上CAN Bus 攻击的具体案例，也就是使用BMS控制开发板做实验的状态

\begin{figure}[!ht]
    \centering
    \includegraphics[width=1\linewidth]{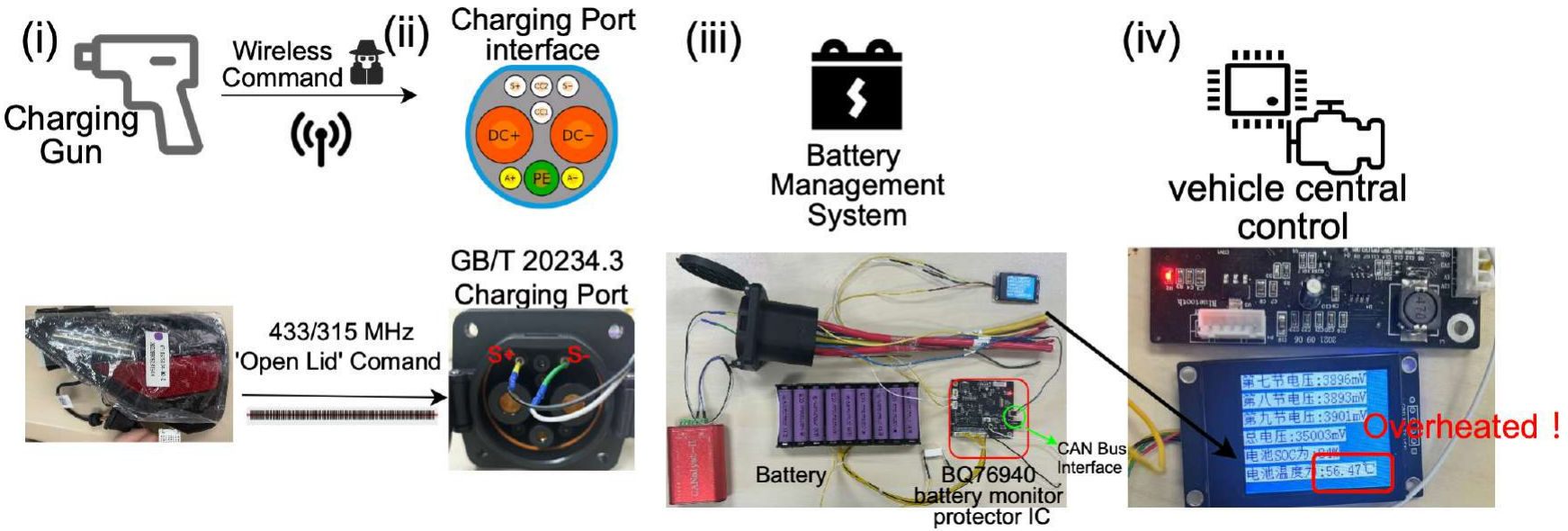}
    \caption{CAN Bus Signal Injection Attack Overview}
    \label{fig:can_injection}
    \vspace{-5px}
\end{figure}

\modify{We explore how CAN bus injection can cause functional compromise of battery management systems (BMS). As shown in Figure~\ref{fig:can_injection}, the attack begins with the replay of the wireless signal used to open the EV charging port, thereby enabling physical access to the S+/S– communication lines defined in the GB/T 20234.3 standard. These lines are commonly connected to CAN bus-based charging controllers in vehicles such as Denza, BYD, XPeng, and Arcfox.}

\modify{To simulate a realistic vulnerability, we built a prototype BMS using the TI BQ76940 battery monitor IC and an STM32F103 MCU, both widely adopted in commercial EVs. The BMS is designed to disconnect charging MOSFETs when the battery temperature exceeds 40 $ ^\circ C $. However, we discovered a stack buffer overflow vulnerability triggered by a specific multi-stage CAN message sequence.}

\modify{Our proof-of-concept payload bypasses this protection by overwriting control register values, resulting in the MOSFET remaining active even under overheating conditions. In our testbed, this led to continued charging until the battery temperature reached 56.47 $ ^\circ C$, significantly exceeding the defined thermal threshold. 
This results in forced charging even under unsafe thermal conditions, with our testbed reaching a battery temperature of 56.47 $ ^\circ C$ in the red box. This demonstrates a critical safety violation caused by code-level flaws reachable via external CAN access.} %\todo{R.3 Added new case study illustrating CAN bus injection leading to BMS protection failure and thermal overrun.}

Although this experiment is conducted in a simulated environment, its architecture reflects real-world systems. For instance, Tesla’s Model S uses a Chargeport ECU that communicates via CAN with internal energy control modules, and lacks strict isolation between the charger-side and internal buses. Our findings suggest that in the absence of proper message authentication or bus isolation, similar injection-based attacks may compromise safety-critical subsystems across various EV platforms.
This case study illustrates that CAN bus injection attacks can lead to persistent and dangerous failures, not just momentary service denial, thus exposing a deeper layer of risk within the EV charging ecosystem.

\subsection{Attack Efficacy}

\begin{table}[!ht]
\centering
\resizebox{0.9\textwidth}{!}{%
\begin{tabular}{@{}cccccc@{}}
\toprule
\textbf{Car Models}                              & \textbf{\begin{tabular}[c]{@{}c@{}}Charging Ports\\ Standard\end{tabular}} & \textbf{DoS Attack}                                & \textbf{Deadlock Attack}                           & \textbf{\begin{tabular}[c]{@{}c@{}}CP PWM\\ Injection Attack\end{tabular}} & \textbf{\begin{tabular}[c]{@{}c@{}}Potential CAN\\ BUS Injection Attack\end{tabular}} \\ \midrule
                                                 & NACS                                                                       & \ding{52}                         & \ding{52}                        & \ding{52}                                                & \ding{52}                                                           \\
                                                 & \cellcolor[HTML]{EFEFEF}SAE J1772                                          & \cellcolor[HTML]{EFEFEF}\ding{52}& \cellcolor[HTML]{EFEFEF}\ding{52}& \cellcolor[HTML]{EFEFEF}\ding{52}                        & \cellcolor[HTML]{EFEFEF}\ding{56}                                    \\
                                                 & CCS I                                                                      & \ding{52}                        & \ding{52}                        & \ding{52}                                                & \ding{56}                                                            \\
                                                 & \cellcolor[HTML]{EFEFEF}GB/T 20234.2                                       & \cellcolor[HTML]{EFEFEF}\ding{52}& \cellcolor[HTML]{EFEFEF}\ding{52}& \cellcolor[HTML]{EFEFEF}\ding{52}                        & \cellcolor[HTML]{EFEFEF}\ding{56}                                    \\
\multirow{-5}{*}{Tesla Model S}                  & GB/T 20234.3                                                               & \ding{52}                        & \ding{52}                        & \ding{52}                                                & \ding{52}                                                           \\ \midrule
                                                 & GB/T 20234.2                                                               & \ding{52}                        & \ding{52}                        & \ding{52}                                                & \ding{56}                                                            \\
                                                 & \cellcolor[HTML]{EFEFEF}GB/T 20234.3                                       & \cellcolor[HTML]{EFEFEF}\ding{52}& \cellcolor[HTML]{EFEFEF}\ding{52}& \cellcolor[HTML]{EFEFEF}\ding{52}                        & \cellcolor[HTML]{EFEFEF}\ding{56}                                    \\
                                                 & IEC 62196                                                                  & \ding{52}                        & \ding{52}                        & \ding{52}                                                & \ding{56}                                                            \\
\multirow{-4}{*}{Tesla Model 3}                  & \cellcolor[HTML]{EFEFEF}CCS II                                             & \cellcolor[HTML]{EFEFEF}\ding{52}& \cellcolor[HTML]{EFEFEF}\ding{56} & \cellcolor[HTML]{EFEFEF}\ding{52}                        & \cellcolor[HTML]{EFEFEF}\ding{56}                                    \\ \midrule
                                                 & NACS                                                                       & \ding{52}                        & \ding{52}                        & \ding{52}                                                & \ding{52}                                                           \\
                                                 & \cellcolor[HTML]{EFEFEF}IEC 62196                                          & \cellcolor[HTML]{EFEFEF}\ding{52}& \cellcolor[HTML]{EFEFEF}\ding{52}& \cellcolor[HTML]{EFEFEF}\ding{52}                        & \cellcolor[HTML]{EFEFEF}\ding{56}                                    \\
\multirow{-3}{*}{Tesla Model Y}                  & CCS II                                                                     & \ding{52}                        & \ding{52}                        & \ding{52}                                                & \ding{56}                                                            \\ \midrule
                                                 & GB/T 20234.2                                                               & \ding{52}                        & \ding{52}                        & \ding{52}                                                & \ding{56}                                                            \\
\multirow{-2}{*}{Volkswagen ID.4}                & \cellcolor[HTML]{EFEFEF}GB/T 20234.3                                       & \cellcolor[HTML]{EFEFEF}\ding{52}& \cellcolor[HTML]{EFEFEF}\ding{52}& \cellcolor[HTML]{EFEFEF}\ding{52}                        & \cellcolor[HTML]{EFEFEF}\ding{52}                                   \\ \midrule
                                                 & GB/T 20234.2                                                               & \ding{52}                        & \ding{52}                        & \ding{52}                                                & \ding{56}                                                            \\
\multirow{-2}{*}{ROEWE RX5}                      & \cellcolor[HTML]{EFEFEF}GB/T 20234.3                                       & \cellcolor[HTML]{EFEFEF}\ding{52}& \cellcolor[HTML]{EFEFEF}\ding{52}& \cellcolor[HTML]{EFEFEF}\ding{52}                        & \cellcolor[HTML]{EFEFEF}\ding{56}                                    \\ \midrule
                                                 & GB/T 20234.2                                                               & \ding{52}                        & \ding{52}                        & \ding{52}                                                & \ding{56}                                                            \\
\multirow{-2}{*}{ARCFOX $\alpha S$ } & \cellcolor[HTML]{EFEFEF}GB/T 20234.3                                       & \cellcolor[HTML]{EFEFEF}\ding{52}& \cellcolor[HTML]{EFEFEF}\ding{52}& \cellcolor[HTML]{EFEFEF}\ding{52}                        & \cellcolor[HTML]{EFEFEF}\ding{56}                                    \\ \midrule
                                                 & IEC 62196                                                                  & \ding{52}                        & \ding{52}                        & \ding{52}                                                & \ding{56}                                                            \\
\multirow{-2}{*}{Li Auto L7}                     & \cellcolor[HTML]{EFEFEF}CCS II                                             & \cellcolor[HTML]{EFEFEF}\ding{52}& \cellcolor[HTML]{EFEFEF}\ding{56} & \cellcolor[HTML]{EFEFEF}\ding{52}                        & \cellcolor[HTML]{EFEFEF}\ding{56}                                    \\ \bottomrule
\end{tabular}%
}
\caption{Effectiveness of Attacks Across Car Models and Charging Standards}
\label{tab:experiment}
\end{table}

We evaluated the efficacy of the proposed attacks across a range of EV models and charging port standards. Table~\ref{tab:experiment} summarizes our findings in a controlled test environment, demonstrating the real-world feasibility of the attacks conducted using \ourwork.

\textbf{DoS Attacks.} The DoS attack proved to be universally applicable across all tested vehicle models and charging standards. By manipulating the CC or CP port impedance, \ourwork consistently triggered the termination of the charging process.

\textbf{CP PWM Signal Injection.} This attack was similarly effective across all configurations, enabling unauthorized control over charging state transitions. In some cases, we observed the ability to start or interrupt charging by replaying or injecting low-duty-cycle PWM signals.

\textbf{Deadlock Attacks.} Unlike the above two, the deadlock attack showed dependency on the physical locking mechanism of the charging interface. Specifically, Tesla Model 3 and Li Auto L7 vehicles equipped with CCS II ports did not exhibit a lock-based constraint and were therefore not susceptible to this attack vector. Other models with electronic locking mechanisms experienced successful deadlock scenarios.

\textbf{CAN Bus Injection.} Finally, the feasibility of CAN bus message injection was contingent on the standard and vehicle design. Only NACS and GB/T 20234.3-based implementations exposed interfaces through which CAN frames could be observed or injected via the charging interface. In contrast, vehicles relying on CCS or SAE J1772 lacked accessible interfaces for direct CAN interaction, preventing this type of exploitation.

Overall, the results confirm that while some attacks are universally effective, others rely on specific hardware designs or protocol implementations. These findings highlight the urgent need to reassess the security assumptions in current EV charging architectures.

\section{Countermeasures} \label{Sec:countermeasures}

\modify{To address vulnerabilities arising from weak authentication in EV charging systems, we propose a more robust, twofold countermeasure that enhances traditional impedance-based methods. Traditionally, EV charging authentication employs a fixed voltage source and a resistor-divider for impedance validation. As shown in Figure~\ref{fig:Inner}, this approach is limited in resisting sophisticated spoofing attacks.}

\begin{figure}[!ht]
    \centering
    \includegraphics[width=0.7\columnwidth]{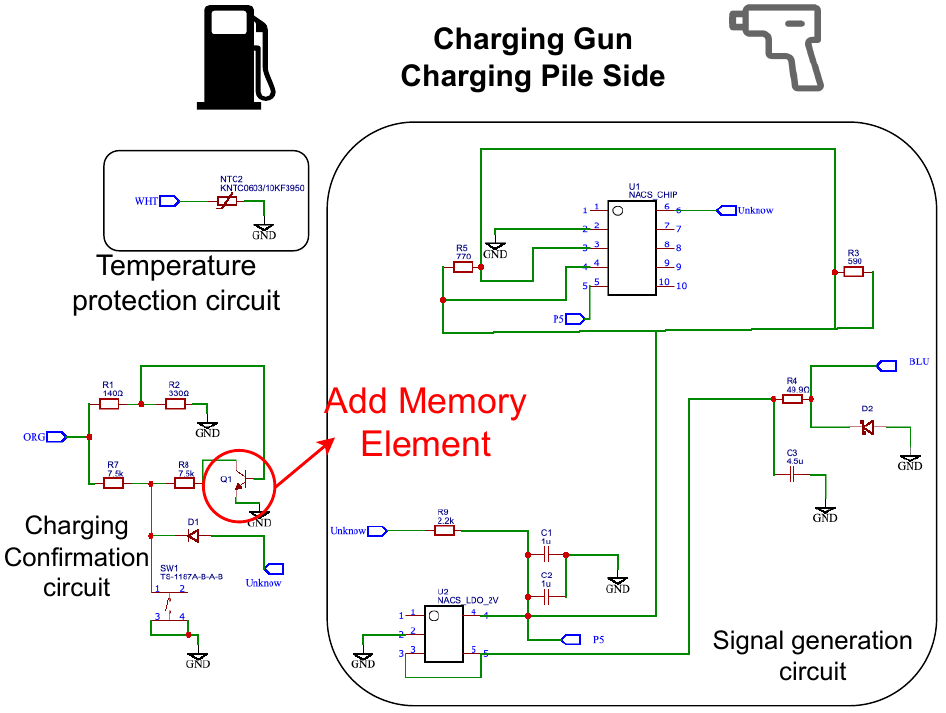}
    \caption{Solution-I: Memory Elements on Charging Gun Side}
    \label{fig:NACS}
    \vspace{-5px}
\end{figure}

Our solution introduces a \textbf{dual-check process} combining the legacy fixed voltage approach with a dynamic signal source. The addition of a variable power source enables comparative analysis of both static and dynamic signal responses, enhancing anomaly detection during authentication.

Security is further reinforced through memory-capable components integrated into the circuit (highlighted in red in Figure~\ref{fig:NACS}), replacing traditional resistors with transistors and capacitors. This is inspired by Tesla’s CC circuit design~\cite{TeslaModelXManual2024,HackerNews2023}. The added complexity alters signal behavior under varying frequencies, thwarting spoofing attempts with fixed resistors.

\begin{figure}[!ht]
    \centering
    \includegraphics[width=0.6\columnwidth]{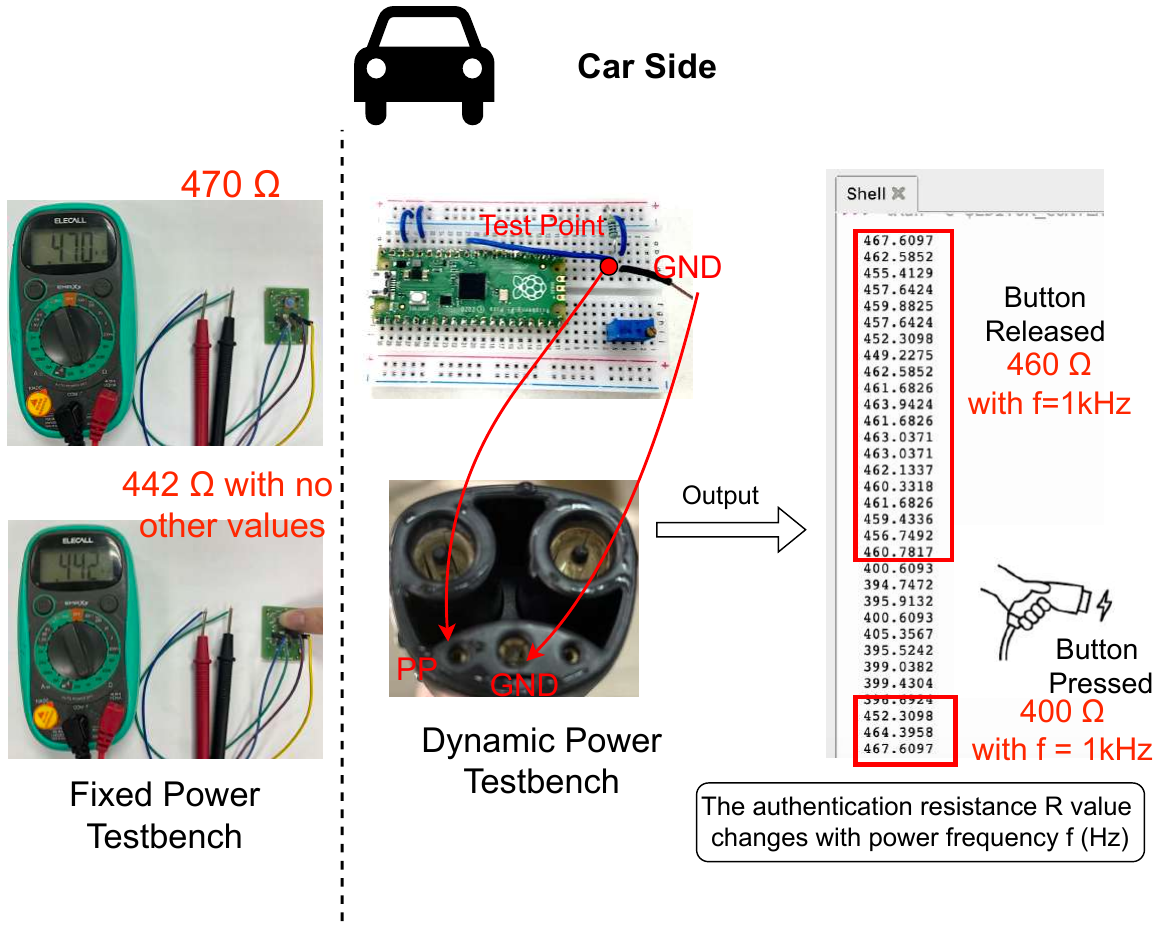}
    \caption{Solution-II: Dynamic Power Source on EV Side} 
    \label{fig:result_nacs}
    \vspace{-5px}
\end{figure}

This design ensures that signals observed through a basic resistor-divider differ significantly from those under dynamic excitation, as shown in Figure~\ref{fig:result_nacs}. Frequency-sensitive components cause impedance shifts: $Z_{\text{transistor}} = j2\pi f L$ and $Z_{\text{capacitor}} = \frac{1}{j2\pi f C}$. A variable signal source modulates these properties, making attacker replication exceedingly difficult.

\modify{By introducing signal variability and component complexity, our method surpasses static resistance checks by adding unpredictability to the validation process and significantly enhancing security. To further strengthen authentication, we propose integrating a lightweight wireless energy transmission module (e.g., RF/NFC) for out-of-band signal generation.} This additional channel enables active tamper detection: any deviation from the expected signal response, under both static and dynamic conditions, can trigger alarms for manual inspection. Combined with anomaly detection in circuit signal characteristics, this integrated approach provides robust protection against spoofing attempts and reinforces the resilience of EV charging systems against sophisticated attacks.

\section{Related Works} \label{Sec:related}

As EVs become increasingly widespread, the security of EVSE has gained critical attention. A growing body of research has investigated vulnerabilities across the charging infrastructure, from remote attacks to physical-layer threats.

\bsub{Network-based EVSE Vulnerabilities.} \modify{Vailoces et al.~\cite{vailoces2023securing} provided a broad analysis of vulnerabilities in EVSE systems, including weak backend authentication and insecure end-to-end communication. Their work outlines several attack scenarios and corresponding countermeasures, primarily at the backend server and network level. In contrast, our work presents the first in-depth analysis and exploitation of state forgery vulnerabilities in the front-end charging protocol between EV and charger, extending the attack surface to include physical and protocol-layer manipulation that can induce denial-of-service, deadlock, and in-vehicle CAN bus injection.} %\todo{R.1 Highlighted our novel contributions beyond [Ano23] and weak-authentication works.}

Johnson et al.~\cite{johnson2022cybersecurity} explored remote spoofing attacks on EV chargers and highlighted weaknesses in authentication flows. While their analysis remains at the simulation level, our study moves beyond theoretical models by building hardware-based attack prototypes and conducting real-world experiments against multiple EV models.

Several studies have investigated publicly accessible EVSE infrastructure. Varriale et al.~\cite{varriale2022risks} and Hille et al.~\cite{hille2018ev} demonstrated that many chargers are discoverable via Shodan and contain outdated firmware, weak credentials, or insecure web interfaces. While these works uncover accessible attack surfaces, our research introduces novel exploitation methods that directly interact with the CP/CC analog signaling to induce misbehavior, representing a more proactive and low-level attack vector.

Liu et al.~\cite{liu2019enhanced} proposed blockchain-based authentication and logging systems to protect EV charging records. While effective for securing data integrity, their approach does not address the physical-layer or analog signaling vulnerabilities we exploit.

% 窃听
% Backer et al. developed wireless eavesdropping tools targeting EV-to-EVSE communication, enabling attackers to steal credentials and influence charging sessions~\cite{baker2019losing}. 
% In contrast, our work focuses on the direct manipulation of charging stations, bypassing the need for credential theft by exploiting inherent weaknesses in the system.Physical vulnerabilities within EV charging components have also garnered attention. Baker et al. analyzed the communication protocols used in EV charging systems, identifying risks at the physical layer, including the exposure of vehicle identification numbers through sniffing techniques \cite{baker2019losing}. While their research underscores the importance of securing physical components, our study extends this focus by examining how physical layer weaknesses enable direct manipulation of charging processes.

\bsub{Physical Layer EVSE Attacks.} \modify{Baker et al.~\cite{baker2019losing} presented credential-stealing attacks via wireless eavesdropping on EV-EVSE communication. By contrast, our work bypasses credential-based assumptions entirely by injecting crafted signals to spoof legitimate EV states, requiring no prior knowledge or access credentials.}

Other physical-layer efforts, such as Kohler et al.'s “Brokenwire” attack~\cite{kohler2022brokenwire} demonstrated large-scale charging disruptions through RF interference. While impactful, their method targets the communication channel as a whole. In contrast, our work precisely targets control signals and impedance states to forge EV charging logic from within the protocol itself.

Kohler et al. examined systemic vulnerabilities across the EV charging infrastructure, showcasing the potential for widespread disruptions that could impact millions of vehicles simultaneously \cite{kohler2022brokenwire}. Their call for comprehensive cybersecurity solutions, incorporating both technical and regulatory frameworks, aligns with the broader industry's need for a multi-layered defense strategy. However, our research zeroes in on physical layer security, presenting new attack vectors that exploit these foundational vulnerabilities in more focused and tangible ways.

Dudek et al.~\cite{dudek2019v2g} released an open-source V2G injector to manipulate XML-based vehicle-to-grid exchanges. Our work complements this by addressing AC/DC handshake and signaling-level weaknesses, particularly those allowing access to the in-vehicle CAN bus via public charging ports.

%区块链
% Additionally, Liu et al. explored the use of blockchain technology to secure EV charging transactions, offering an innovative solution for enhancing cybersecurity by leveraging decentralized, tamper-resistant systems~\cite{liu2019enhanced}. Their approach effectively safeguards charging data from spoofing attacks, though it remains primarily focused on data integrity rather than the broader spectrum of physical and protocol-based attacks, which are the central focus of our research.

% 组织防御
% Wasumwa et al. emphasized the dynamic nature of cybersecurity threats and the necessity of adaptive security strategies that go beyond technical solutions to include policy and regulatory frameworks~\cite{wasumwa2023safeguarding}. While their work provides a holistic approach to EV charger security, our research narrows the focus to the technical vulnerabilities that directly enable system manipulation, contributing new insights into attack methodologies that previous studies have overlooked.

\bsub{Practical EVSE Exploitation.}The Southwest Research Institute (SwRI)~\cite{swri2024evcharging, swri2024evcybersecurity} explored the J1772 protocol and proposed a zero-trust model for charging networks. However, their limited focus and lack of empirical validation leave key protocol-layer attack vectors unexplored. Our work addresses this gap by systematically evaluating real-world AC and DC charging standards (e.g., GB/T 20234.2/.3, NACS), revealing authentication flaws that permit attacker-controlled state transitions.

Wasumwa et al.~\cite{wasumwa2023safeguarding} highlighted the importance of policy and adaptive defense frameworks. While their perspective is valuable for large-scale ecosystem protection, our research focuses specifically on the technical root causes that enable real-time EVSE manipulation, contributing hands-on methodologies for exploitation and defense.

\bsub{Signal Injection Attacks.}
Recent research has uncovered emerging attack vectors leveraging signal injection. Wang et al.~\cite{wang2022ghosttalk} demonstrated how inaudible power-line noise can affect EV charging control logic. Similarly, laser-based voice injection attacks~\cite{shi2024laser,sugawara2020light} show that unconventional signals can be used to trigger unintended behaviors. \modify{Inspired by these, our work contributes the \textbf{first practical demonstration of signal injection into the charging interface}, exploiting physical signal-level authentication flaws to control EV states directly.}

\section{Discussion} \label{Sec:discussion}

\subsection{Ethical Considerations}
All experiments in this study were conducted on EVs owned by the authors. No third-party vehicles or public users were involved, ensuring that our research caused no unintended disruption or harm.
We followed ethical research guidelines throughout the entire process, including full compliance with relevant data protection laws, safety protocols, and usage consent to contribute positively to the EV security ecosystem through transparent and responsible disclosure.

\modify{To ensure that our work aligns with the broader interests of the community, we engaged with safety experts. We also verified that all test signals introduced during experimentation, such as PWM pulses and CAN frames, were confined within hardware safety thresholds and did not pose any risk of physical damage, data leakage, or irreversible changes to vehicle systems.} %\todo{R.6 Revised paper structure and corrected typos.}

\subsection{Responsible Disclosure}

To ensure the ethical handling of vulnerabilities identified in EV charging systems, we completed a coordinated responsible disclosure process before publication. The initial disclosure was conducted through the organizers of the GEEKCON competition\footnote{GEEKCON, formerly known as GeekPwn, is a premier security event in China where researchers publicly demonstrate cutting-edge vulnerabilities and novel attack surfaces.}, where our work was selected and showcased as a live demonstration to the security community and participating vendors.

\modify{Following this initial presentation, we formally reported the vulnerabilities to the China National Vulnerability Database (NVDB), specifically through the China Automotive Vulnerability Database (CAVD). Our submissions have been assigned four official identifiers. In addition to national database disclosures, we have also submitted four CVE requests through a recognized CNA partner, and these applications are currently under review.}

\modify{The vulnerabilities were disclosed to major stakeholders across the EV ecosystem, including but not limited to Seres, Denza, Zeekr, BYD, XPeng, Arcfox, and Dongfeng Motor. Several vendors acknowledged receipt of the disclosure and expressed intent to investigate or mitigate the reported issues. A detailed timeline of the disclosure process, including vendors and impacts, is included in Appendices.} % \todo{R.5 Added responsible disclosure process details, including vendor notifications, NVDB/CVE submission status, and timeline.}

\subsection{Implications}

Our research uncovers critical weaknesses in the authentication mechanisms of publicly deployed EV charging systems, revealing the ease with which adversaries can exploit physical-layer vulnerabilities to manipulate charging states and vehicle behavior. Using our \ourwork attack suite, we demonstrated the ability to remotely interfere with charging processes and simulate discharge signals through PWM injection.

These findings highlight a broader systemic issue: the authentication protocols embedded in many EV charging infrastructures remain overly simplistic and insufficiently protected against signal-level spoofing. Given the adoption of international standards across diverse vendors, the risks identified here are not isolated to a single manufacturer but span a wide range of EV models and charging pile deployments.

The most immediate risk is the potential for vehicles to be unlawfully immobilized or mischarged at public stations, causing severe disruption to users. More concerning, however, is the long-term exposure these weaknesses create, particularly in environments where vehicles are integrated into smart transportation grids or used for Vehicle-to-Grid (V2G) interactions.

%\modify{Beyond EVs, our findings underscore a fundamental challenge for any cyber-physical system: weak or static authentication at the hardware interface can undermine the entire trust model, regardless of higher-layer protections. These issues have direct implications for other infrastructure domains, including smart energy meters, industrial IoT, and autonomous transportation.}

To mitigate such risks, manufacturers must adopt more robust, multi-layered authentication approaches—including physical signal verification, real-time anomaly detection, and cryptographic validation. Regulators should also update certification requirements and standards to mandate stronger safeguards at both the protocol and hardware levels.

Finally, our findings point to future research directions. These include extending our threat model to private home charging systems, analyzing backward compatibility concerns for proposed countermeasures, and applying similar techniques to evaluate authentication protocols in other critical embedded systems.

\subsection{Limitations} 

As with any practical security research, our work has certain limitations defined by the scope of our experiments and the constraints of real-world systems.

First, the \ourwork attack suite was designed to target widely deployed public EV charging piles, and may not apply to newer systems that adopt proprietary or advanced authentication protocols. Notably, we did not test against CHAdeMO~\cite{nakanishi2019chademo} or ChaoJi, limiting coverage across all global standards.

Second, although our experiments were conducted in realistic charging environments, they were still controlled to ensure safety. Some extreme behaviors, such as disabling hardware-level protections or simulating malicious firmware inside EVs, were beyond our scope due to ethical constraints. For instance, when experimenting with CP line PWM injection, internal Battery Management System (BMS) safeguards prevented us from inducing high-current damage, demonstrating the effectiveness of built-in safety designs, but also limiting attack depth.

Lastly, we focused on protocol-level and analog signal-layer attacks. Future work could explore cross-layer attacks that integrate physical-layer spoofing with protocol fuzzing, or conduct broader empirical studies involving international vendors and private charging facilities.

% \vspace{-10px}

% The limitations of our work primarily stem from two key factors: the design goals of our attack and the scope of our experimental testing. First, the \ourwork attack suite may not be universally applicable to all existing charging systems, particularly those equipped with more advanced or non-standard security features. This means that some systems, especially those adopting proprietary or next-generation protocols, might not be vulnerable to the same attacks demonstrated in our research. Notably, we did not conduct experiments on CHAdeMO~\cite{nakanishi2019chademo} or the second-generation ChaoJi charging standard, which limits the scope of our findings.

% From a methodological perspective, our experiments were conducted in controlled environments, which may not fully replicate all real-world conditions. As such, the behavior of the attacks in diverse operational scenarios, such as under varying network loads or environmental conditions, remains untested. For instance, while we explored CP line signal injection to manipulate charging current, the built-in protection mechanisms of the vehicle's Battery Management System (BMS) against overheating prevented us from directly causing battery damage through the injection of high-duty-cycle PWM signals. This suggests that while our attacks can disrupt operations, they may not always result in immediate catastrophic damage, depending on the vehicle’s internal safeguards.

\section{Conclusion} \label{Sec:conclusion}

Our research reveals significant vulnerabilities in the authentication mechanisms of EV charging systems, specifically highlighting weak points in widely adopted protocols. Through the use of \ourwork, we successfully demonstrate the feasibility of remote manipulation of charging operations, showcasing attack vectors such as signal injection and manipulation of the 
% Control Pilot (CP) and Charging Confirmation (CC) ports. 
CP and CC ports. 
These vulnerabilities expose public charging infrastructures to potential threats, where attackers could exploit weak authentication processes to disrupt or immobilize vehicles, posing both safety and security risks.

Our findings suggest that current authentication protocols across various charging standards—including GB/T 20234, IEC, SAE J1772, NACS, and CCS—are inadequate in defending against sophisticated spoofing attacks, particularly in systems that rely on static resistance-based authentication mechanisms. The potential for injecting malicious signals, such as CAN bus messages, further underscores the critical need to reevaluate the security of charging infrastructure as EV adoption accelerates globally. In addition to exposing vulnerabilities, we also propose countermeasures to mitigate these risks. These include enhancing authentication protocols by integrating dynamic power, memory electric elements, and multi-layer security checks.

\section*{Appendices} \label{Sec:Appendices}

\modify{The table~\ref{tab:CAVD} lists four assigned vulnerability IDs affecting major EV vendors, along with brief impact summaries and their corresponding disclosure timelines.} 

\begin{table}[!ht]
\centering
\resizebox{\textwidth}{!}{%
\begin{tabular}{@{}cccc@{}}
\toprule
\textbf{Vuln. ID}    & \textbf{Affected Vendor(s)}                                                                              & \textbf{Impact Summary}                                                                                                                                                                   & \textbf{Disclosure Timeline}                                                                                 \\ \midrule
\textbf{NVDB-CAVD-2025478822} & \begin{tabular}[c]{@{}c@{}}Seres, Denza, Zeekr, BYD, \\ XPeng, Arcfox \\ and Dongfeng Motor\end{tabular} & \begin{tabular}[c]{@{}c@{}}Weak resistance-based authentication allows an \\ attacker to simulate invalid CC states, causing\\  the charger to reject charging (DoS attack).\end{tabular} & \begin{tabular}[c]{@{}c@{}}Reported: 2025-03-18  \\ Acknowledged: 2025-05-22 \\  Fixed: pending\end{tabular} \\
\rowcolor[HTML]{EFEFEF} 
\textbf{NVDB-CAVD-2025018034} & \begin{tabular}[c]{@{}c@{}}Seres, Denza, Zeekr, BYD, \\ XPeng,Arcfox \\ and Dongfeng Motor\end{tabular}  & \begin{tabular}[c]{@{}c@{}}Forged CC resistance locks the charging gun, \\ preventing removal (deadlock attack).\end{tabular}                                                             & \begin{tabular}[c]{@{}c@{}}Reported: 2025-03-18  \\ Acknowledged: 2025-05-22 \\  Fixed: pending\end{tabular} \\
\textbf{NVDB-CAVD-2025864575} & \begin{tabular}[c]{@{}c@{}}Seres, Denza, Zeekr, \\ BYD, Arcfox\end{tabular}                              & \begin{tabular}[c]{@{}c@{}}Malicious CC values trigger discharge mode, \\ draining the EV battery.\end{tabular}                                                                           & \begin{tabular}[c]{@{}c@{}}Reported: 2025-03-18  \\ Acknowledged: 2025-05-22 \\  Fixed: pending\end{tabular} \\
\rowcolor[HTML]{EFEFEF} 
\textbf{NVDB-CAVD-2025820938} & \begin{tabular}[c]{@{}c@{}}Denza,  BYD, XPeng\\ Arcfox\end{tabular}                                      & \begin{tabular}[c]{@{}c@{}}Bypassing CC2 check enables CAN injection, \\ allowing remote control of charging.\end{tabular}                                                                & \begin{tabular}[c]{@{}c@{}}Reported: 2025-03-18  \\ Acknowledged: 2025-04-10 \\  Fixed: pending\end{tabular} \\ \bottomrule
\end{tabular}%
}
\caption{Vulnerability Disclosure Summary}
\label{tab:CAVD}
\end{table}

%%%% 8. BILBIOGRAPHY %%%%
%\bibliographystyle{alpha}
\bibliographystyle{IEEEtran}
\bibliography{ref}
%%%% NOTES
% - Download abbrev3.bib and crypto.bib from https://cryptobib.di.ens.fr/
% - Use bilbio.bib for additional references not in the cryptobib database.
%   If possible, take them from DBLP.

\end{document}